\documentclass[5p]{elsarticle}

\biboptions{numbers, sort&compress}
\usepackage[resetlabels, labeled]{multibib}
\newcites{P}{Primary Studies}

\makeatletter
\def\NAT@spacechar{}
\makeatother

\usepackage{multirow}

\usepackage{threeparttable}
\usepackage{booktabs}
\usepackage{amsmath}
\usepackage{graphicx}
\usepackage{subfigure}

\usepackage{float}
\usepackage{stfloats}
\usepackage{url}

\journal{Information \& Software Technology}

\begin{document}
	
\begin{frontmatter}
	
\title{A Systematic Review of Unsupervised Learning Techniques for Software Defect Prediction}

\author[myfirstaddress,myfourthaddress]{Ning Li}

\author[mysecondaryaddress]{Martin Shepperd}
\cortext[mycorrespondingauthor]{Corresponding author: Martin Shepperd}
\ead{martin.shepperd@brunel.ac.uk}

\author[mythirdaddress]{Yuchen Guo}

\address[myfirstaddress]{\leftline {School of Computer Science, Northwestern Polytechnical University, Xi'an, 710072, China}}
\address[mysecondaryaddress]{\leftline {Brunel University London, Uxbridge, UB8 3PH, United Kingdom}}
\address[mythirdaddress]{\leftline {Department of Computer Science and Technology, Xi'an Jiaotong University, Xi'an, 710049, China}}
\address[myfourthaddress]{\leftline {Key Laboratory of Big Data Storage and Management, Ministry of Industry and Information Technology, Xi'an, 710072, China}}

\begin{abstract}
	\textbf{Background}: Unsupervised machine learners have been increasingly applied to software defect prediction. It is an approach that may be valuable for software practitioners because it reduces the need for labeled training data. \\
	\textbf{Objective}: Investigate the use and performance of unsupervised learning techniques in software defect prediction.\\
	\textbf{Method}: We conducted a systematic literature review that identified 49 studies containing 2456 individual experimental results, which satisfied our inclusion criteria published between January 2000 and March 2018. In order to compare prediction performance across these studies in a consistent way, we (re-)computed the confusion matrices and employed the Matthews Correlation Coefficient (MCC) as our main performance measure.\\ 
	\textbf{Results}: Our meta-analysis shows that unsupervised models are comparable with supervised models for both within-project and cross-project prediction. Among the 14 families of unsupervised model, Fuzzy CMeans (FCM) and Fuzzy SOMs (FSOMs) perform best.  In addition, where we were able to check, we found that almost 11\% (262/2456) of published results (contained in 16 papers) were internally inconsistent and a further 33\% (823/2456) provided insufficient details for us to check.\\
	\textbf{Conclusion}:  Although many factors impact the performance of a classifier, e.g., dataset characteristics, broadly speaking, unsupervised classifiers do not seem to perform worse than the supervised classifiers in our review.  However, we note a worrying prevalence of (i) demonstrably erroneous experimental results, (ii) undemanding benchmarks and (iii) incomplete reporting. We therefore encourage researchers to be comprehensive in their reporting.
\end{abstract}

\begin{keyword}
	Unsupervised learning; Software defect prediction; Machine learning; Systematic review; Meta-analysis.
\end{keyword}
	
\end{frontmatter}

\section{Introduction}\label{sec:Intro}

\noindent	
Various software defect prediction models have been proposed to improve the quality of software over the past few decades \cite{fenton1999critique}.  An increasingly popular approach is to use machine learning \cite{catal2009systematic, hall2012systematic}.  These approaches can be divided into supervised methods where the training data requires labels, typically faulty or not, and unsupervised methods where the data do not need to be labelled.  Supervised prediction models predominate.  However, in practice it is often difficult to collect defect classification labels to train a Supervised Defect Prediction (SDP) model \cite{nam2015clami}.  As a consequence in recent years, Unsupervised Defect Prediction (UnSDP) models have begun to attract attention. 

The main aim of our systematic review is to provide software practitioners and researchers with guidance for software defect prediction, particularly regarding whether the use of unsupervised prediction models is a viable option. We analyse 49 UnSDP primary studies that satisfy our inclusion criteria.  From these primary studies, we investigate which unsupervised learning algorithms were deployed and the relative predictive performance of supervised and unsupervised models. 

Our systematic review makes the following contributions:
\begin{enumerate}
	\item Identification of a set of 49 primary studies related to UnSDP published between January 2000 and March 2018. These cover a wide range (14) of unsupervised prediction technique families including six different cluster labelling techniques.
	\item Data extraction and regularisation from the 49 studies satisfying our inclusion criteria. We obtain 2456 individual experimental results from these studies.		 
	\item A meta-analysis to compare the relative performance of unsupervised and supervised learners from two perspectives (i) the specific learning algorithm and (ii) the dataset. Based on the analysis results we make suggestions for machine learning-based, software defect prediction for practitioners.
	\item A bibliometric analysis that considers research and publishing trends, along with the quality of reported experimental results.
\end{enumerate}

The remainder of this paper is organised as follows: Section~\ref{sec:Review} describes our systematic review methodology including the research questions, study selection criteria, data extraction and synthesis (process, prediction performance measures and raw data summary). Next, Section~\ref{sec:Bibmetrics} presents current research trends and quality issues: both incomplete and inconsistent reporting of results. Section~\ref{sec:Perform} shows our meta-analysis results. Section~\ref{sec:Threats} discusses the threats to validity, followed by conclusions and discussion of actionable results in Section~\ref{sec:Conc}.		

\section{Review Methodology}\label{sec:Review}
\noindent
Our systematic review follows the guidelines of a systematic review approach for software engineering presented in \cite{Kitc15}.  First, we clarify our research questions, then search and identify relevant primary studies, next we synthesise results from the selected primary studies, finally we explore and answer our research questions through meta-analysis.

\subsection{Research Questions}

\begin{itemize}
	\item RQ1: What are the publication trends in unsupervised software defect prediction research?
	\item RQ2: What is the quality of the experimental reporting (completeness and consistency)?
	\item RQ3: What kinds of unsupervised learning research experiments are conducted?
	\item RQ4: What is the difference between unsupervised and supervised software defect predictive performance?
	\item RQ5: Which unsupervised prediction models or model families perform better? 
	\item RQ6: What is the impact of dataset characteristics on predictive performance?	
\end{itemize}

\noindent
The first three research questions will principally be of interest to software engineering researchers.  The remaining three questions will be of interest to both practitioners and researchers.

\subsection{Search Process}
\noindent
We used five search engines to include the papers published between January 2000 and March 2018.  The start date was aligned with the search period of the widely cited systematic review by Hall et al.~\cite{hall2012systematic}.  We undertook our search on 7th March, 2018.  The search engines include the ISI Web of Science, ACM Digital Library, IEEE Xplore, ScienceDirect and SpringerLink.  Although there are small variants in the five search engines, our key search string is: 

\begin{verbatim}	
	("fault prediction" OR "defect prediction" 
	OR "bug prediction" OR "error prediction")
	AND ("unsupervised" OR "unlabel*" 
	OR "cluster*") 
	AND ("software")
\end{verbatim}

Table~\ref{tbl:review} presents the results of our paper search and selection process. This results in 49 papers selected from an initial 1360 papers. We list these 49 primary papers at the end of this paper where P$i$ denotes the i\textsuperscript{th} primary study.  Note that experimental results in \cite{yan2016self} and  \cite{boucher2016using} are included in \citep[P][]{yan2017automated} and \citep[P][]{boucher2018software} respectively, thus we only took \citep[P][]{yan2017automated} and \citep[P][]{boucher2018software} as our primary studies.  

\begin{table}[ht]
	\caption{Systematic review paper search and selection process}		
	\centering
	\footnotesize 
	\label{tbl:review}
	\begin{tabular}{|p{4.3cm}|p{1.5cm}|p{1.3cm}|}\hline
		Search and selection process & \# of added / excluded papers & \# of remaining papers  \\\hline
		Search of five academic search engines  & +1360  & 1360 \\\hline
		Exclude duplicate studies & -346 & 1014  \\\hline
		Exclude irrelevant studies based on title and abstract & -943 & 71   \\\hline
		Forward and backward chain using Google Scholar & +88 & 159 \\\hline
		Exclude papers that use defect labels for training data or unsupervised techniques are used only for pre-processing & -96 & 63   \\
		\hline
		Exclude papers without real-world data experiments & -6 &   57  \\\hline
		Exclude papers with insufficiently detailed or duplicated results & -8 & 49 \\\hline
	\end{tabular}	
\end{table}

\subsection{Inclusion Criteria}

\noindent
When determining whether a paper or experimental result should be included or not, the following criteria were applied. 

\begin{enumerate}
	\item Written in English.
	\item Full content must be available.
	\item Published between January 2000 - March 2018.
	\item Uses real data (not simulations).
	\item Applies at least one unsupervised method to software defect-prone module prediction. 
	\item Includes new software defect prediction experiments.  We ignore re-analysis of previously published experiments.
	\item Includes the best prediction performance results per dataset and learner (since some experiments have a primary focus on some other aspects of machine learning e.g., feature selection, parameter tuning).
	\item Reports the results in sufficient detail to enable meta-analysis.
\end{enumerate}


\subsection{Data Extraction and Synthesis} \label{subsec:dataprocess}

\noindent	
We extracted data related to our research questions from each of the 49 papers, and organised the qualitative and quantitative data into a raw data file (see our Mendeley dataset \cite{Li2020Data}) . Each paper contains from 1 to 751 ( median = 12) experimental results, yielding a total of 2456 individual results, which involve 128 distinct software project defect datasets (from NASA, ISM, AEEEM, PROMISE, etc.) and 25 prediction model families (14 unsupervised learners and 11 supervised learners). 

\subsubsection{Data Extraction and Synthesis Process}
\noindent
From each paper we extracted the following:
\begin{itemize}
	\setlength{\itemsep}{0pt}
	\setlength{\parskip}{0pt}
	\item Title
	\item Year
	\item Journal/conference
	\item `Predatory' publisher? (Y $ | $ N)
	\item Count of results reported in paper
	\item Count of inconsistent results reported in paper 
	\item Parameter tuning in SDP? (Yes $ | $ Default $ | $ ?) 
	\item SDP references\footnote{Here we seek to understand the researchers' approach to SDP benchmark through their citations of related work and published results.} (SDPRefs$\_ $OrigResults  $ | $  SDPRefs  $ | $  SDPNoRefs  $ | $  OnlyUnSDP)
\end{itemize}

\noindent	
Then, from within each paper, we extracted for each experimental result:
\begin{itemize}
	\setlength{\itemsep}{0pt}
	\setlength{\parskip}{0pt}
	\item Prediction method name (e.g., DTJ48)
	\item Project name trained on (e.g., PC4)
	\item Project name tested on (e.g., PC4)
	\item Prediction type (within-project $ | $ cross-project)
	\item No. of input metrics (count $|$ NA)
	\item Dataset family (e.g., NASA)
	\item Dateset fault rate (\%)
	\item Was cross validation used? (Y $|$ N $ | $ ?)
	\item Was error checking possible? (Y $ | $ N)
	\item Inconsistent results? (Y $ | $ N $ | $ ?)
	\item Error reason description (text)
	\item Learning type (Supervised $|$ Unsupervised)
	\item Clustering method? (Y $|$ N $|$ NA)
	\item Machine learning family (e.g., Un-NN)
	\item Machine learning technique (e.g., KM) 
	\item Prediction results (including TP, TN, FP, FN, etc.)
\end{itemize}

For full details refer to the review protocol at our Mendeley dataset \cite{Li2020Data}. To ensure data quality, we undertook pre-processing (synthesising across studies) including name unification (project name, method name, response variable name, etc.), confusion matrix (re-)computation\footnote{We parenthesise the `re' of computation to convey that for some papers we are constructing the confusion matrix \emph{ab initio}.  In other situations there is already an explicit matrix but we re-construct it from other data provided, although potentially by different means from that presented in the original paper.  For brevity, in the remainder of our paper we simply state `re-computation'. }(see Table~\ref{tbl:matrix}) and data quality checking with R scripts. We describe the details of confusion matrix re-computation and data quality checks in Section~\ref{subsec:quality}.

\subsubsection{Prediction Performance Measures} \label{subsec:perfMetric}
\noindent	
For data classification, the confusion matrix (Table~\ref{tbl:matrix}) is the fundamental descriptor from which the majority of performance indicators may be derived.  Although ideally all primary studies would report consistent performance indicators, in practice, a wide range of indicators are used such as accuracy, precision, recall, the F-measure\footnote{The F-measure that researchers use is the balanced F score and is often referred to as F1.  Other weightings are possible but in practice seldom used \cite{Powe11}.}, the G-measure and so forth.  Consequently, we reconstruct the confusion matrix wherever possible.  Unfortunately, there remain about 33\% (823/2456) of the experimental results for which this was not possible, due to incomplete reporting as discussed in Section~\ref{subsec:LowQual}.  

The performance indicators used by our set of 49 primary studies are summarised below. To further complicate matters, note that different studies may use different names for the same measure  \cite{shepperd2014researcher, bowes2014dconfusion}.

\begin{table}[h]
	\caption{Confusion Matrix}		
	\centering
	\footnotesize
	\label{tbl:matrix}
	\begin{tabular}{|c|c|c|}\hline
		&  Observed defective &  Observed defect free \\\hline
		Predicted &  True Positive & False Positive \\
		defective &  (TP) & (FP)  \\\hline
		Predicted &  False Negative & True Negative \\
		defect free & (FN) & (TN)  \\\hline	
	\end{tabular}	
\end{table}

\begin{enumerate}
	\item FPR (False Positive Rate): $ \frac{FP}{TN+FP} $ 
	\item FNR (False Negative Rate): $ \frac{FN}{TP+FN} $
	\item ER (Error Rate): $ \frac{FP+FN}{TP+TN+FP+FN} $
	\item Recall: $ \frac{TP}{TP+FN} $ 
	\item Accuracy: $ \frac{TP+TN}{TP+TN+FP+FN} $
	\item Precision: $ \frac{TP}{TP+FP} $
	\item F1: $ \frac{2 \times Recall \times Precision}{Recall+Precision} $
	\item MCC: $ \frac{TP \times TN - FP \times FN}{\sqrt{(TP+FP)(TP+FN)(TN+FP)(TN+FN)}} $ \cite{baldi2000assessing}
	\item AUC: Area Under the Curve (AUC) ROC chart \cite{fawcett2006introduction}.
	\item Popt/ACC: Effort-aware prediction performance \cite{kamei2013large}.
	\item G-Mean:$\sqrt{ \frac{TP}{TP+FN} *  \frac{TN}{TN+FP}}$ (\citep[P][]{boucher2018software}).
	\item G-Measure: $ \frac{2 \times Recall \times TNR}{Recall+TNR} $ . (\citep[P][]{fan2017utility})
	\item Balance: $ 1- \frac{\sqrt{(0-FPR)^{2} + (1-Recall)^{2} }}{\sqrt{2}} $. (\citep[P][]{singh2014efficient})
	\item Purity: Percentage of the most-dominated category (fault prone or
	not fault prone) in the cluster. \citep[P][]{zhong2004analyzing}
	\item MAE: Mean absolute error between prediction and observation (\citep[P][]{chug2013software}).
	\item MeanAIC: The mean of Akaike's Information Criterion (AIC) (\citep[P][]{iwata2012clustering}).
\end{enumerate}

Despite being widely used, F1 and AUC are known to be potentially problematic \cite{Hand09,Powe11,Flac15}.  F1 focuses on positive classes and ignores negative classes.
This is acceptable for information retrieval type problems (since the number of say irrelevant documents, correctly not retrieved can be essentially unbounded).  However, this is not the case for defect prediction because the precision of negative classes (non-defective) is also of concern to software developers. Knowing that a component is non-defective is important.  Thus the application domain of defect prediction is unsuitable for F1.

The AUC metric compares algorithms without classification threshold.  So it is suitable to compare classifier \emph{families} in a theoretical sense but not in a practical sense.  The reason being, in practice a deployed classifier must have a particular classification threshold.  So unless a classifier strictly dominates another (i.e., for all possible threshold values) we cannot use AUC to prefer one classifier to another.

In contrast, the Matthews Correlation Coefficient (MCC) \cite{baldi2000assessing} is based on all four quadrants of the confusion matrix, which gives a better summary of the performance of classification algorithms.  It is also known by statisticians as the $\phi$ coefficient, originally proposed by Karl Pearson.  MCC is easier to interpret as correlation coefficient since it takes a value in the interval [-1, 1], with 1 showing a perfect classifier, -1 showing a perverse classifier, and 0 showing that the prediction is uncorrelated with the ground truth.   A more detailed comparison between MCC, F1 and AUC can be found in our another work \cite{song2018comprehensive} (see Section 3.1 Classification Performance Measures).  Therefore, we prefer to use MCC to assess classification performance in this paper. Note, however, that MCC is undefined if any of the quantities TP + FN, TP + FP, TN + FP, or TN + FN are zero \cite{baldi2000assessing}. 

\subsubsection{Data Summary} \label{subsec:rawdata}
\noindent
From the 49 studies we obtain 2456 individual experimental results.  Table~\ref{tbl:statClass} presents a summary of the categorical attributes.  Over half of the results are from supervised learners which are generally deployed as comparators to the unsupervised methods. Within-project (as opposed to cross-project) classification is the dominant approach.  

The remaining categories relate to primary study quality which we explore in Section~\ref{subsec:quality}.

\begin{table}[ht]
	\caption{Summary of key categorical attributes for individual results} 
	\label{tbl:statClass}
	\centering
	\footnotesize
	\begin{tabular}{|l|l|}	\hline
		Attribute & Count \\\hline
		Learning Type & Unsupervised: 947; Supervised: 1509  \\
		Prediction type & WithinPrj: 1840; CrossPrj: 616 \\ 
		\hline
		Cross validation & Yes: 2210; Unknown: 246 \\ 
		`Predatory' publisher& Yes: 280;  No: 2176 \\ 
		Inconsistent data & Yes: 262; No: 1371; Unknown: 823 \\ \hline		
	\end{tabular}
\end{table}

Likewise, Table~\ref{tbl:statNumber} summarises the numerical attributes where \#NA is the number of unavailable results. From our data extraction and synthesis, 262 inconsistent results were identified (see Section~\ref{subsec:quality}).  Therefore, Table~\ref{tbl:statNumber} only lists a summary of the valid --- in the sense of being internally consistent --- 2194 experimental results after removing 262 inconsistent ones.  Quite striking is the diversity of values, e.g., the fault rate $d$ ranges from 0.4\% to over 93\%. The AUC ranges from 0.31 to 0.948 (recall that any value below 0.5 suggests a classifier that is predicting worse than by chance). Note that \#NA of MCC is larger than 823 unchecked results, which is caused by zero in the denominator of its definition (see Section~\ref{subsec:perfMetric}). F1 is similarly impacted.  For the fault rate, \#NA is due to unreported values in private (RSDIMU, CSAS, MIS and Embedded used in 20 experiments) or modified public datasets (KC3\_314 and JM1\_8916 are public NASA data, but are inconsistent with the original datasets and are used in a further 4 experiments). However, the overall unavailability of fault rate data is low, considering the total number of datasets and fortunately has little effect on our meta-analysis.

\begin{table}[ht]
	\caption{Summary of 2194 consistent experimental results (including 823 unchecked results)} 
	\label{tbl:statNumber}
	\centering
	\footnotesize
	\begin{tabular}{|l|r|r|r|r|r|}
		\hline
		Statistic & Fault rate ($d$) & MCC & F1 & AUC & Popt \\ 
		\hline
		Min 	& 0.004 & -0.524 	& 0.016  & 0.310 	& 0.276 \\ 
		Q1 		& 0.122 & 0.245 	& 0.381  & 0.610 	& 0.583 \\ 
		Median  & 0.204 & 0.384 	& 0.531  & 0.670 	& 0.726 \\ 
		Mean    & 0.237 & 0.360 	& 0.519  & 0.674 	& 0.697 \\ 
		Q3 		& 0.310 & 0.494 	& 0.654  & 0.750 	& 0.815 \\ 
		Max    	& 0.936 & 0.931 	& 0.978  & 0.948 	& 0.948 \\ 
		\#NA    & 24     & 952    	& 945     & 1734 	& 1843 \\ 
		\hline
	\end{tabular}
\end{table}

\subsection{Quality Factors}\label{subsec:quality}
\noindent
In order to understand the quality of the different studies within our analysis, we consider two classes of explanatory factors. External factors are publication venue and type, whereas internal factors address experimental design and consistency of results.

First we explore publication venue.  Our search involves a wide range of journal or conference publications.  Note that the number of experimental results in these primary studies varies greatly from 1 to 751 thus some papers have more propensity for error. We classify the 49 selected papers into two groups according to whether they are found in the so-called `predatory' publisher lists \cite{perlin2017predatory}. The list aims to identify standalone publishers with questionable approaches to peer review rigour.

We used three lists: for publisher\footnote{https://beallslist.weebly.com/}, for standalone journals\footnote{https://beallslist.weebly.com/standalone-journals.html} and for conferences\footnote{https://libguides.caltech.edu/c.php?g=512665\&p=3503029}.  Note that these lists have been questioned \cite{Berg15} so we recognise the classification is imperfect.  

Second, we consider internal experiment quality.  For this we check for the use of cross-validation \cite{kohavi1995study}, as an indicator for experimental quality.   More importantly, we check the internal consistency of results. Hence we recompute the confusion matrix, or construct it if not provided.  This is possible for any one of the 10 combinations of performance measure reporting listed below, i.e., we solve for X.  From this we can compute MCC.

For brevity we just present the details for Cases 9 and 10 (N.B.~Cases 1 to 8 can be found in \cite{bowes2014dconfusion}).  Here, $d$ is the proportion of defective modules, $a$ is accuracy, $r$ is recall and $p$ is precision.  Other definitions for FPR, FNR, ER, Accuracy, etc. are given in Section~\ref{subsec:perfMetric}. Note that we use other measures in preference to $d$, since real $d$ might be slightly divergent from reported $d$ in the original study due to pre-processing or cross validation and differences between folds. Therefore, we only use $d$ when we cannot recompute the confusion matrix for Cases~1-4. Furthermore, we normalize TP, TN, FP and FN with $ TP+TN+FP+FN=1 $ when preprocessing raw data.

\begin{enumerate}
	\item FPR, FNR, ER
	\item FPR, FNR, Accuracy
	\item FPR, Recall, Precision
	\item Recall, Precision, Accuracy
	\item Defect Percentage(d), FPR, FNR
	\item Defect Percentage(d), FPR, ER
	\item Defect Percentage(d), FPR, Recall
	\item Defect Percentage(d), Recall, Precision
	\item Defect Percentage(d), Accuracy, F1:  \\
	$ TP = \frac{(1-a)*F1}{2*(1-F1)} $, 	
	$ TN = a -TP $, 
	$ FN = d -TP $, 
	$ FP = 1-TP-TN-FN $ .
	\item Defect Percentage(d), Precision, F1: \\
	$ TP = \frac{p*d*F1}{2*p-F1} $, 	
	$ FN = d -TP $, 
	$ FP = TP *(1-p)/p$,\\
	$ TN = 1 - TP-FN-FP $. 	
\end{enumerate}

As part of our re-computation, we also check the correctness and consistency of the experimental results.  If the reported or recomputed result satisfies one of the following six inconsistency rules, we label it as problematic data and remove it from our data analysis. In total, we removed 262 results (Rule~1: 171; Rule~2: 7; Rule~3: 60; Rule~4: 3; Rule~5: 19; Rule~6: 2) out of 2456 experimental results.  

Rule~1. The recomputed performance values (including TP, F1,  MCC, d, etc.) are out of their correct ranges (for example: $TP \notin [0,1]$, $ F1 \notin [0,1] $,  $MCC \notin [-1,1] $,  $d \notin [0,1] $ ).

Rule~2. The recomputed $ d $ is 0 (we treat 0 as problematic data since all experiments for defect prediction should include defective modules).

Rule~3. Compare the recomputed $d$ with the original reported defect percentage, we treat them as problematic data if the difference is greater than 0.1 (i.e., we allow for some rounding errors).

Rule~4. Compare our recomputed results with the original ones where available.  If the difference of a measure is greater than the rounding error range, we label them as problematic data. Here, the rounding error intervals are computed by adding $\pm0.05$ to the original data.  Note, compared to the rounding error tolerance of 0.01 used by \cite{bowes2014dconfusion}, we apply a wider range of 0.05. 


Rule~5. Special values from Cases 1--10. Here, we use Case1 as example. For a given FPR, FNR, ER in Case1, the confusion matrix is recomputed. We derive the formulae: $ d = \frac{ER-FPR}{FNR-FPR} $, $ FN= d*FNR $,  $ FP= (1-d)*FNR$, $TP=d-FN $, $TN=1-TP-FP-FN $. If the numerator is not zero and the denominator is zero, then it is a problematic result.  In this case, if $ER \neq FPR $ and $FNR=FPR$ (e.g., FNR=0.24, FPR=0.24, ER=0.13), it will be labeled as problematic or inconsistent. A complete explanation for all of 10 cases can be found in our  Mendeley dataset (see Section~\ref{subsec:dataprocess}).

Rule~6. Obvious problematic experimental data is reported by the primary study. For example, the confusion matrix in \citep[P][]{sandhu2010k} is inconsistent with their dataset. In \citep[P][]{sandhu2010k}, their Table X shows that there are 134 bugs (BUGs=true) in their JEdit dataset, however the following clustering prediction result Table XI and Table XII show that there are only 61 bugs in that dataset (Table XIII: TP=46, FN=15).

\section{Bibliometric analysis} \label{sec:Bibmetrics}

\subsection{RQ1: What are the publication trends in unsupervised software defect prediction research?} \label{subsec:PubTrends}
\begin{figure}[htb]
	\begin{center}
		\includegraphics[width=.9\linewidth]{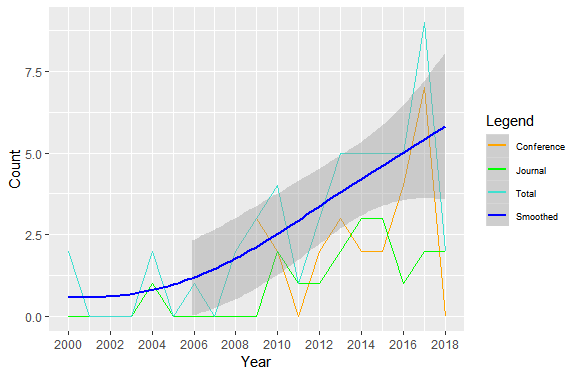}
		\caption{Unsupervised software defect prediction papers (2000-2018).  The broad blue line shows total publications using a loess smoother and 95\% confidence limits where these are non-negative, i.e., 2005-onwards.  N.B.~2018 is incomplete.}
		\label{Fig:PubTrends}
	\end{center}
\end{figure}

\noindent
Recall that our search has identified 49 primary studies that have applied unsupervised learning to the task of software defect prediction.  These are published over the period of 2000-2018\footnote{Note that year 2018 is incomplete.} and are summarised in Table~\ref{Tab:PapersByYear} and also by Figure~\ref{Fig:PubTrends}.   From this we can see that conference papers tend to dominate and that there has been pronounced growth in the overall number of papers published in recent years as indicated by the smoother in Figure~\ref{Fig:PubTrends}.

\begin{table}[h]
	\caption{Research papers published by type and year} 
	\label{Tab:PapersByYear}
	\centering
	\begin{tabular}{|r|r|r|r|}
		\hline
		& Conf & Journal & Total \\ 
		\hline
		2000 & 2 & 0 & 2 \\ 
		2004 & 1 & 1 & 2 \\ 
		2006 & 1 & 0 & 1 \\ 
		2008 & 2 & 0 & 2 \\ 
		2009 & 3 & 0 & 3 \\ 
		2010 & 2 & 2 & 4 \\ 
		2011 & 0 & 1 & 1 \\ 
		2012 & 2 & 1 & 3 \\ 
		2013 & 3 & 2 & 5 \\ 
		2014 & 2 & 3 & 5 \\ 
		2015 & 2 & 3 & 5 \\ 
		2016 & 4 & 1 & 5 \\ 
		2017 & 7 & 2 & 9 \\ 
		2018 & 0 & 2 & 2 \\ 
		\hline
		Total & 31 & 18 & 49 \\ 
		\hline
	\end{tabular}
\end{table}

\subsection{RQ2: What is the quality of the experimental reporting (completeness and consistency)?}\label{Subsec:quality}

\noindent
The 49 primary studies contain a total of 2456 individual experimental results.   This ranges from 1 to 751 results with the median number being 12 results contained in a single paper.   Problematic results are analyzed in Section~\ref{Subsec:quality}, so here we just provide an overall summary of the remaining 2194 non-erroneous results in Table~\ref{Tab:ResErrCt}.

\begin{table}[htb]
	\caption{Counts and proportions of experimental results}
	\begin{center}
		\begin{tabular}{|l|r|r|}
			\hline
			Result & Count & \% of total \\
			\hline
			Inconsistent results & 262 & \ 10.67\% \\
			Non-error results & 2194 & 89.33\% \\
			\hline
			(Non-error) Cannot check & 823 & 33.51\% \\
			(Non-error) Can check - ok & 1371 & 55.82\% \\
			\hline
			Total & 2456 & 100\% \\
			\hline
		\end{tabular}
	\end{center}
	\label{Tab:ResErrCt}
\end{table}

\subsubsection{`Predatory' publishing}
\begin{figure}[ht]
	\begin{center}
		\includegraphics[width=.9\linewidth]{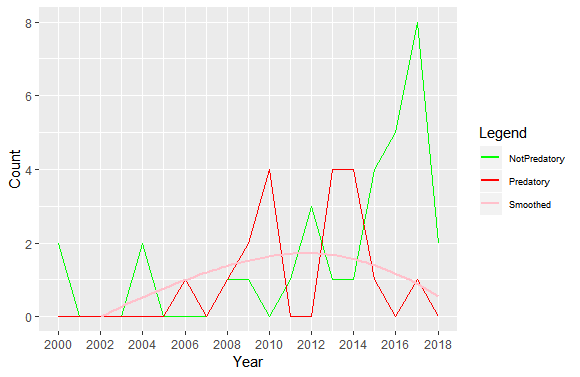}
		\caption{Unsupervised software defect prediction papers (2000-2018) by venue.  The broad pink line shows `predatory' publications using a loess smoother.  N.B.~2018 is incomplete.}
		\label{Fig:PubQualityTrends}
	\end{center}
\end{figure}

\noindent
Next we consider the prevalence of papers published in so-called `predatory' journals and conferences. Table~ \ref{Tab:PredPubByYear} reveals that 18 out of the 49 papers (i.e., approximately 37\%) were from venues considered to be `predatory' publishers.  This was a higher proportion than we had expected.  This phenomenon appears to have started in the late 2000s and is possibly declining since 2014 (see also Figure~\ref{Fig:PubQualityTrends}).  This could be a reaction to the adverse publicity such publishers are gaining.

\begin{table}[ht]
	\caption{Papers published by year and by venue, i.e., `predatory' or `non-predatory' publisher} 
	\label{Tab:PredPubByYear}
	\centering
	\begin{tabular}{|r|r|r|r|}
		\hline
		& No & Yes & Sum \\ 
		\hline
		2000 & 2 & 0 & 2 \\ 
		2004 & 2 & 0 & 2 \\ 
		2006 & 0 & 1 & 1 \\ 
		2008 & 1 & 1 & 2 \\ 
		2009 & 1 & 2 & 3 \\ 
		2010 & 0 & 4 & 4 \\ 
		2011 & 1 & 0 & 1 \\ 
		2012 & 3 & 0 & 3 \\ 
		2013 & 1 & 4 & 5 \\ 
		2014 & 1 & 4 & 5 \\ 
		2015 & 4 & 1 & 5 \\ 
		2016 & 5 & 0 & 5 \\ 
		2017 & 8 & 1 & 9 \\ 
		2018 & 2 & 0 & 2 \\ 
		\hline
		Sum & 31 & 18 & 49 \\ 
		\hline
	\end{tabular}
\end{table}

\subsubsection{Problems with Incomplete Reporting}\label{subsec:LowQual}

Next we consider \emph{potential} issues which may arise from incomplete reporting of experimental design and results. First, we examine whether a study explicitly reports if a cross-validation procedure was used or not.  We were surprised to find that 25 out of 49 studies did \emph{not} report this information.  This is important because although unsupervised learners need not initially be trained with data that have class labels (i.e., defective or not), the learner must still be configured to classify and hence be evaluated on unseen data.  Of course some, or even all, of the other studies may have applied cross-validation techniques, however this \emph{ought} to be confirmed by the authors in the first instance.

In terms of results, as described in Section~\ref{subsec:quality} we endeavoured to apply basic consistency checks to the raw results.  Unfortunately this was not possible for over 30\% of the results (823/2456) listed in Table~\ref{Tab:ResErrCt}.  Given, as we discuss in the following section, that we uncovered 262 (262/1633 $ \approx $ 16\%) instances of problematic data in the results we are able to check, this is a little disturbing.

\subsubsection{Errors and inconsistent results} \label{subsec:Inconsistent}

\begin{figure}[ht]
	\centering
	\includegraphics[width=\linewidth]{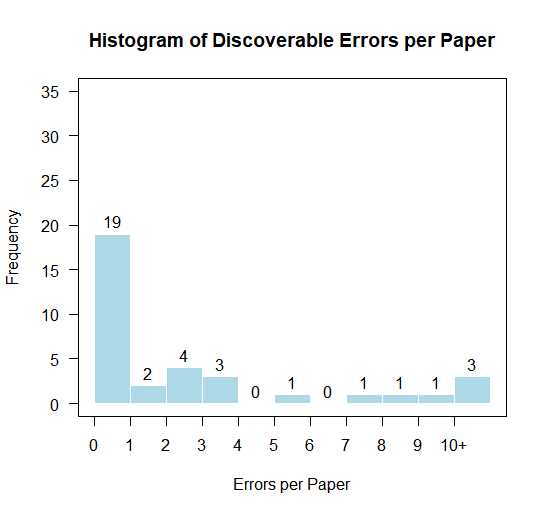}
	\caption{Histogram of discovered errors (inconsistencies) per paper}
	\label{Fig:ErrsPerPaper}
\end{figure}

As previously indicated, we were able to apply consistency checking to approximately 66\% (1633 out of 2456) of all the results in our systematic review.  Figure~\ref{Fig:ErrsPerPaper} shows the distribution of errors (or inconsistencies) per paper. There are 14 papers that are excluded due to incomplete reporting.  Note also that in order to improve the visual qualities of the histogram, the largest bin is 10+ and it contains a single extreme outlier of 165 inconsistencies and two studies including 19 and 30 inconsistencies respectively.

Next, we consider whether there is any relationship between publication venue and study errors.  For example, one might expect that `non-predatory' publishers deploy more rigorous peer review, or that `predatory' publishers are less diligent in this regard.  To validate such an expectation, we examine the odds ratio \cite{grissom2012effect} which is defined as:

\[ OR = \frac{Odds_{pc}}{Odds_{notpc}} =  \frac{P(T|pc)/P(not T|pc)}{P(T|not pc)/P(not T|not pc)} \]

\noindent
where $Odds_{pc}$ is the odds that a member of Population $pc$ will fall into Category $T$, similarly $Odds_{notpc}$ is the odds that a member not from Population $pc$ will fall into Category $T$.  

When calculating the odds ratio (from Table~\ref{Tab:ErrorsPredPub}) the likelihood of containing an error is $ 1.21  $(with 95\% confidence limits $[0.31, 4.73]$) between `non-predatory' and `predatory' publishers.  This is close to unity and so suggests little effect at the study level.  However, a similar analysis for individual experimental results (from Table~\ref{Tab:ErrorsDataNum}) reveals a different picture of $0.02$ (with 95\% confidence limits $[0.02, 0.04]$).  This suggests individual results from a `non-predatory' paper clearly are less likely to contain errors than a `predatory' paper, however, this is skewed by a small number of papers that contain very high numbers of results (and inconsistencies).

\begin{table}[ht]
	\caption{Paper contains consistency errors and `predatory' publisher} 
	\label{Tab:ErrorsPredPub}
	\centering
	\begin{tabular}{|r|r|r|r|}
		\hline
		Predatory? & Error  & Ok & Unchecked \\ 
		\hline
		No  & 10 & 11 & 10  \\
		`Predatory' & 6 & 8 & 4 \\  		
		\hline
	\end{tabular}
\end{table}

\begin{table}[ht]
	\caption{Predatory publisher and Error Data} 
	\label{Tab:ErrorsDataNum}
	\centering
	\begin{tabular}{|r|r|r|r|}
		\hline
		Predatory? & Error & Ok & Unchecked \\ 
		\hline
		No 	& 88  & 1307 & 781\\  
		`Predatory'  & 174  & 64 & 42\\
		\hline
	\end{tabular}
\end{table}

\subsubsection{Use of comparator benchmarks} \label{subsubsec:benchmarks}

When assessing prediction methods such as UnSDP it is usual to provide some kind of comparator or benchmark.  This is particularly helpful when a new or improved learning method is proposed.  However, there is a risk that most of the research `energy' is invested in the new method, as opposed to the benchmark.  As Michie et al.~\cite{Mich94} remarked 25 years ago, we need to be vigilant about the problems of comparing `pet' algorithms in which researchers are expert with others with which they are less familiar.  They also note the more general danger of selecting comparator benchmarks that are not state of the art.  How does this apply to our systematic review?  It potentially raises a bias most obviously with RQ4 that compares UnSDP and SDP predictive performance.

\begin{table}[ht]
\caption{Use of Supervised Learner Benchmarks and the Literature (by paper)}
\begin{center}
	\begin{tabular}{|p{1.8cm}|p{5cm}|p{1cm}|}
		\hline
		\textbf{SDPRef} & \textbf{Description} & \textbf{Study Count} \\
		\hline
		SDPRefs + OrigResults & Supervised methods refer to some literature, and they use the results reported in those literature & 4 \\
		SDPRefs & Supervised methods refer to some literature, but do not use the reported results & 16 \\
		SDPNoRefs & Supervised methods used but no reference to literature (5 out of 8 papers used very common supervised learners, e.g., Na\"{\i}ve Bayes, Random Forest, Logistic Regression) & 8 \\
		Only UnSDP & No supervised comparator models & 21 \\
		\hline
		Total &  & 49 \\
		\hline
	\end{tabular}
\end{center}
\label{Tab:UseOfBenchmarks}
\end{table}

Unfortunately we cannot know researchers' intentions and making judgements as to when a method is a `pet' requires a great deal of subjectivity.  However, we can at least examine how papers use the literature with regard to SDPs.  Table~\ref{Tab:UseOfBenchmarks} shows that of the 28 papers that use SDPs as a comparator for their UnSDPs, 71\% (20/28) cite related articles, although only four use previous prediction results.  In contrast 8/28 papers use supervised methods without citation suggesting that common methods such as logistic regression are seen as basic default methods that do not warrant discussion or tuning (see Section~\ref{subsec:finemcc} for an analysis of hyper-parameter tuning).

\subsection{RQ3: What kinds of unsupervised learning research experiments are conducted?}


\noindent
For a more detailed analysis of unsupervised learning techniques, we categorise them into clustering and non-clustering techniques as presented in Table~\ref{tbl:UnSvSDP}.  We further divide these unsupervised prediction techniques into 14 families and 21 sub-families. 

\begin{table*}[htbp]
	\caption{Unsupervised Software Defect Prediction Techniques}		
	\centering
	\footnotesize
	\label{tbl:UnSvSDP}
	\begin{threeparttable}
		\begin{tabular}{|p{1.7cm}|p{4.7cm}|p{1cm}|p{6.5cm}|p{1.5cm}|}\hline
			\textbf{Clustering?} & \textbf{MethodFamily } & \textbf{Sub. Abbr} &\textbf{SubFamily Approaches} &  \textbf{Related Study} \\	\hline
			\multirow{2}{*}{\textbf{Clustering}}	
			& \multirow{3}{4.5cm}{\textbf{k-Partition (KPart)}: Based on the distance of data points, assign them into k exclusive partitions or groups, where each partition represents a cluster.} 
			& KM   & K-means: Each data point is assigned into the cluster according to the distance with the centroid of a cluster, which is the mean of all points within the cluster. The initial number of clusters is assigned manually. & \multirow{3}{1.55cm} {\citep[P][]{catal2009software, c2009clustering, zhong2004analyzing, z2004unsupervised, bishnu2012software, b2011application, singh2016comparative, varade2013hyper, coelho2014applying, chug2013software, gupta2013estimating, yang2008software, yang2006software, kaur2014comparative, sandhu2010k,  singh2014efficient, kaur2010clustering}, \citep[P][]{zhang2016cross},\citep[P][]{aleem2015benchmarking}, \citep[P][]{abaei2015empirical},\citep[P][]{singh2017classification}, \citep[P][]{chang2017novel}, \citep[P][]{yan2017automated, boucher2018software,sasidharan2014hyper}}  \\ \cline{3-5}
			&   & XM  & X-means: An extended K-Means which tries to automatically determine the number of clusters.  
			&\citep[P][]{catal2009software},\citep[P][]{catal2010metrics}, \citep[P][]{park2014software}\\ \cline{3-5}
			&   & KMD & K-medoids: A clustering algorithm related to the k-means, while using an actual point in the cluster as centroid to clustering.  &\citep[P][]{b2011application}, \citep[P][]{zhang2016cross}\\  \cline{2-5}
			& \textbf{Hierarchical (HC)}: Grouping data into a hierarchy of clusters. & HC   & Hierarchical clustering: Can be either agglomerative (bottom-up) or divisive (top-down). &\citep[P][]{chug2013software}, \citep[P][]{gupta2013estimating} \\ \cline{2-5}
			& \textbf{Density (Density)}: Discover non-spherical clusters by finding core objects with dense neighborhoods. & DBC   & Density based clustering: Connects core objects and their neighborhoods to form dense regions as clusters. &\citep[P][]{bishnu2012software},\citep[P][]{chug2013software}, \citep[P][]{sandhu2010density}\\ \cline{2-5}
			& \multirow{2}{4.5cm}{\textbf{Neural Network (NN)}: Clustering based on Neural Network.} 
			& NGas  & {Neural-Gas clustering: A competitive learning technique with SoftMax learning rule.} & \multirow{2}{1.5cm}{\citep[P][]{zhong2004analyzing},\citep[P][]{z2004unsupervised}, \citep[P][]{abaei2015empirical},\citep[P][]{zhang2016cross}}\\ \cline{3-5}
			& 	& SOMs   & Self-Organizing Maps: Produces a low-dimensional  representation of the input space. 
			&\citep[P][]{iwata2013error,abaei2013fault,marian2015software,czibula2016novel,iwata2012clustering}, \citep[P][]{abaei2015empirical},  \citep[P][]{boucher2017predicting},\citep[P][]{boucher2018software}\\ \cline{2-5}
			& \multirow{3}{4.5cm}{\textbf{Fuzzy Logic (Fuzzy) }: Each data point can belong to two or more clusters in fuzzy clustering.} 
			& FCM & Fuzzy C-Means: Assigns membership to each data point corresponding to each cluster center on the basis of distance between the center and itself.  &\multirow{3}{1.5cm}{\citep[P][]{c2009clustering},\citep[P][]{gupta2013estimating}, \citep[P][]{yang2006software}, \citep[P][]{abaei2015increasing},\citep[P][]{zhang2016cross}} \\ \cline{3-5}
			&   & FSubC  &  Fuzzy subtractive clustering: Uses fuzzy inferences based on rules generated by subtractive clustering. &\citep[P][]{yuan2000application},\citep[P][]{sidhu2010subtractive} \\  \cline{3-5}	
			&   & FSOMs & Fuzzy Self-Organizing Maps: Combines SOMs with the concept of fuzziness in fuzzy clustering.   &\citep[P][]{czibula2016novel} \\ \cline{2-5}	
			& {\textbf{Spectral Clustering (SC)}: Make use of the eigenvalues of the similarity matrix to reduce dimensionality before clustering.} 
			& SC & Partitions a dataset based on the connectivity between its nodes in a graph with spectral clustering. & \citep[P][]{zhang2016cross},\citep[P][]{chang2017novel}\\ \cline{2-5}
			& \textbf{Expectation maximization clustering (EM)}: Get probabilities of cluster memberships based on probability distributions. & EM  & EM clustering: Finds the maximum likelihood by iteratively alternating between expectation step and maximization step.  &\citep[P][]{coelho2014applying},\citep[P][]{yang2006software}, \citep[P][]{park2014software},\citep[P][]{guo2000software}, \citep[P][]{chang2017novel} \\ \cline{2-5}
			& \textbf{Optimization (Opt)}: Taking clustering as an optimization problem, such as apply Particle Swarm Optimization to solve.  & PSO & Particle Swarm Optimization clustering: Find the optimal clusters by iteratively  improving a candidate solution with regard to a given measure. &\citep[P][]{coelho2014applying} \\ \cline{2-5}
			& \textbf{Affinity Propagation (AP)}: Finds exemplars, members of the input set that are representative of clusters. & AP  & Affinity Propagation clustering: Obtains the data similarities, and iteratively exchanges the real-valued messages between data points to clustering.  &\citep[P][]{yang2008software},\citep[P][]{yang2008Metrics}\\ \cline{2-5}
			& \multirow{3}{4.5cm}{\textbf{Self Learning (SL) }: Clustering by comparing each metric value with the corresponding median or average value.} 
			& CLA & Clustering and labelling: Compute instances violation degree based on the relation between metric values and corresponding median values, then clustering by the violation or consistency information. &\multirow{3}{1.5cm}{\citep[P][]{yang2016defect},\citep[P][]{n2015clami} } \\ \cline{3-5}
			&   & CLAMI  & Based on CLA, add two extra steps: Metric selection and instance selection based on violation degree. &\citep[P][]{yang2016defect},\citep[P][]{n2015clami}, \citep[P][]{yan2017automated}\\  \cline{3-5}	
			&   & ACL & Average Clustering and labelling: Clustering by metrics of instances violation score (MIVS), which is computed based on the relation between each metric value and their average metric value.  &\citep[P][]{yang2016defect} \\ \cline{2-5}	
			& \textbf{Clustering Ensemble (CLEM)}: Ensemble multiple clustering solutions. & CLEM & Clustering Ensemble: Uses multiple algorithms to clustering, then combines the different clustering solutions and produces a single cluster. &\citep[P][]{coelho2014applying}  \\ \hline
			\multirow{2}{1.7cm}{\textbf{No Clustering}}	& \textbf{Threshold (THD)$^{1} $}: Threshold for selected metric. It might be determined by experiences, rules, calculations, or machine learning, etc. &THD  & Threshold: Use some metric thresholds to classify instances directly. e.g., given a metric threshold vector, if any metric of an instance exceeds the corresponding threshold, it will be defect-prone.  &\citep[P][]{c2009clustering},\citep[P][]{zhong2004analyzing}, \citep[P][]{abaei2013fault},\citep[P][]{n2015clami}, 
			\citep[P][]{catal2013fault},\citep[P][]{abaei2015empirical},
			\citep[P][]{boucher2018software}  \\ \cline{2-5}
			&\textbf{Expert (EXP)}: Determine by experience. &EXP  & Expert determine the labels of each instance directly by their experience.  &\citep[P][]{n2015clami} \\ \cline{2-5}
			&\textbf{Ranking (MR)}: Metric Ranking  & MR  & Metric Ranking: For some change metrics in just-in-time prediction, using 1/Metric to rank instances in ascending order.  &\citep[P][]{yang2016effort},\citep[P][]{fu2017revisiting}, \citep[P][]{liu2017code,yan2017file,fan2017utility,huang2017supervised}, \citep[P][]{chen2018multi} \\ \hline
		\end{tabular}
		\begin{tablenotes}
			\item[1] For `Threshold', if the threshold is calculated with defective labels in some method, we will mark it as supervised method, e.g. STHD-ROC in  \citep[P][]{boucher2018software} .
		\end{tablenotes}
	\end{threeparttable}
\end{table*}

Clustering-based prediction approaches dominate.  They typically include two phases: clustering and labelling. Firstly, all instances in a dataset are clustered into different groups, then each group is labeled as defective or not.  In addition, there are also a few non-clustering unsupervised prediction techniques, such as using thresholds that label instances directly.   

Labelling is a necessary step in unsupervised clustering defect prediction. Table~\ref{tbl:labelling} summarises all labelling approaches located by our review. This shows considerable diversity.  Generic thresholds, as opposed to Pareto or distribution methods, are the most popular. However, practitioners have to consider what is the most suitable labelling technique when employing some particular learners, such as Clustering and Labelling (CLA, see Table~\ref{tbl:UnSvSDP}).

\begin{table*}[t]
	\caption{Labelling techniques for different clusters}		
	\centering
	\footnotesize
	\label{tbl:labelling}
	\begin{threeparttable}
		\begin{tabular}{|p{3.5cm}|p{8cm}|p{4.5cm}|} 
			\hline
			\textbf{Approaches} &  \textbf{Description} & \textbf{Paper}  \\	\hline
			Expert  &  Decided cluster label by expert & \citep[P][]{zhong2004analyzing},\citep[P][]{z2004unsupervised}, \citep[P][]{n2015clami} \\ \hline
			Threshold  & Decided cluster label by particular metric threshold values & \citep[P][]{c2009clustering}, \citep[P][]{abaei2015empirical}\\ \hline
			Distribution& According to the number of nodes in a cluster, scattered (smaller) is defect-prone; concentrative (larger) is defect-free. &\citep[P][]{yang2008software},\citep[P][]{park2014software},\citep[P][]{guo2000software}, \citep[P][]{yuan2000application},\citep[P][]{yang2008Metrics}\\ \hline
			Majority vote  & Majority voting by the most three similar clustering centres. &\citep[P][]{abaei2015increasing} \\ \hline
			Top half & Half of the top clusters are defect-prone, e.g., rank clusters according to violation degree in descending order. &\citep[P][]{zhang2016cross},\citep[P][]{n2015clami},\citep[P][]{yan2017file}, \citep[P][]{yan2017automated} \\ \hline
			Supervised learning   & Use supervised models to predict cluster labels.  &  \citep[P][]{boucher2018software} \\ \hline
		\end{tabular}
		\begin{tablenotes}
			\item[1] `Expert' and `Threshold' are used for cluster label decision in Table~\ref{tbl:labelling}, and for instance label decision in Table~\ref{tbl:UnSvSDP}.
		\end{tablenotes}
	\end{threeparttable}
\end{table*}

\section{Analysis of Unsupervised Defect Prediction Performance} \label{sec:Perform} 

\noindent
This section addresses research questions RQ4--RQ6. In RQ4, we conduct a meta-analysis by vote-counting.  This is not an ideal approach to meta-analysis, but necessary in order to avoid losing important primary studies which employ a range of response measures (performance measures) and experimental designs.  Meta-analysis by vote-counting is a simple, though less powerful, procedure than directly using effect size, to compare the performance of two groups.  It is only recommended when there are problems with a direct approach \cite{Bush09}.  

Our meta-analysis is hindered in that some studies only report AUC. So even though the experiments are informative, they would not contribute to the meta-analysis if we choose F1 to make comparisons.  Essentially vote-counting does not directly use effect size estimates from each primary study.  Although at its simplest it proceeds by comparing the number of `positive' studies with `negative' studies\footnote{Despite the recent emergence of software defect studies that only vote count studies where there is a statistically `significant' effect sometimes referred to as a win-tie-loss procedure, this is in error and studies should always be included \cite{Most96,Bush09}.}, more sophisticated vote-counting methods are available (see Bushman and Wang \cite{Bush09} for an overview).  We use the Hedges and Olkins method for unequal sample sizes \cite[pp.~47--74]{Hedges1985statistical} in order to compare UnSDP and SDP.

\begin{table}[ht]
	\caption{The volume of data involved in analysis} 
	\label{Tab:NumAnalysis}
	\centering
	\footnotesize
	\begin{threeparttable}
		\begin{tabular}{|p{1.8cm}|p{2.5cm}|p{0.8cm}|p{0.7cm}|p{0.8cm}|}
			\hline
			\multicolumn{2}{|p{4.8cm}|}{Description}  & $ Study $ & $ Expt^{1} $ & $ Prj^{1} $ \\ \hline
			\multicolumn{2}{|p{4.8cm}|}{Total meta-data} & 49 & 2456 & 128 \\ \hline
			\multirow{2}{1.8cm} {Removing 262 inconsistent results} & be counted if some result is consistent or incomplete$ ^2 $ & 47 & 2194  & 125 \\ \cline{2-5}
			& not be counted as long as a result is inconsistent $ ^2 $  & $ 33^3 $ & 2194 & 89 \\ \hline
			\multicolumn{2}{|p{4.8cm}|}{Include both unsupervised and supervised learning (used in RQ4 vote-counting)} & $ 26^{4} $ expts: 2052 & - & 110 expts: 2128   \\ \hline 								
			\multicolumn{2}{|p{4.8cm}|}{High quality MCC: MCC is not NA and Predatory $=$ No (used in RQ4 MCC comparison and RQ6)} & 22 & 1178 & 85 \\ \hline		
			\multicolumn{2}{|p{4.8cm}|}{High quality Popt: Popt is not NA and Predatory $=$ No (used in RQ4 Popt comparison)}  & 5 & 351 & 16 \\\hline	 
			\multicolumn{2}{|p{4.8cm}|}{unsupervised learning in high quality MCC of within-project prediction (used in RQ5 MCC comparison)}  & 22 & 426 & 85 \\ \hline	
			\multicolumn{2}{|p{4.8cm}|}{unsupervised learning in high quality Popt of within-project prediction (used in RQ5 Popt comparison)}  & 5 & 107 & 16  \\ 
			\hline				
		\end{tabular}
		\begin{tablenotes}
			\item[1] Expt = count of experimental results; Prj = count of software projects.
			\item[2] One study could contain both inconsistent and consistent results. When we use "be counted if some result is consistent or incomplete", then Expt will be increased by 1, while when using "not be counted as long as a result is inconsistent", Expt will not be increased. Similarly with Prj.
			\item[3] There are 16 studies that include inconsistent experiments according to Figure~\ref{Fig:ErrsPerPaper}, thus only 33 studies could be used.
			\item[4] In total, there are 28 studies that include both UnSDP and SDP model, but all of the SDP experiment results in these two studies are inconsistent, so we removed these two papers.	
		\end{tablenotes}
	\end{threeparttable}
\end{table}

We also carry out a finer grained meta-analysis. Table~\ref{Tab:NumAnalysis} describes the number of studies, experimental results and projects (or datasets) in our meta-analysis.  Note that in the RQ4 vote-counting comparison, 26 studies including 2052 results are involved when we take study as a voting unit, and 110 projects including 2128 results are employed when project is taken as the voting unit.

\subsection{RQ4: What is the difference between unsupervised and supervised software defect predictive performance?} 
\label{subsec:UnSDPvSDP}

\noindent
In this section, we compare UnSDP and SDP models at a coarse (Section~\ref{subsec:vote-counting}) and finer (Section~\ref{subsec:finemcc}) level respectively. All 1306 supervised experimental results are taken from our 26 primary studies which include both unsupervised and supervised models.

\subsubsection{Comparison by vote-counting with multiple measures} \label{subsec:vote-counting}

We carry out vote-counting comparison in this section.  Note the possibility that some conclusions might be unstable depending on whether making the voting unit a  project or study (paper).  For example, this has been a factor for studies \citep[P][]{yang2016effort}, \citep[P][]{fu2017revisiting}.  So we carry out the vote-counting analysis from both perspectives: (i) by study (each study can vote once) and (ii) dataset (a vote per project). Based on the method of Hedges and Olkin \cite{Hedges1985statistical}, we carry out the vote-counting as follows.

\begin{enumerate}
	\item Identify voting units which must include \emph{both} unsupervised and supervised prediction models to enable comparability. Voting unit type can be study or project. Depending on the choice of voting unit, all eligible studies or projects are identified. Each eligible study or project will vote once to determine whether UnSDP is better than SDP.
	
	\item Determine the sample size (number of experiments) for each paper or project, and calculate mean, standard deviation of response variable for each group (UnSDP and SDP). The priority of response variable is: MCC, F1, AUC, Popt, Accuracy, ER, Recall, Precision, GMean, ACC, GMeasure, Balance, Purity, MAE and MeanAIC. For example, if MCC and F1 are available in primary study $ P1 $,  MCC will be used to vote for $ P1 $. This priority is determined by the number of all available results for each response variable among 2194 consistent experiment results. 
	
	\item Determine $X$ for each voting unit, where $ X $ is the sign of ES (Effect Size based on the mean difference between two groups). If $ \bar{Y}_{i}^{U} - \bar{Y}_{i}^{S} > 0 $, $ X =1$ otherwise,  $ X=0 $, where $ \bar{Y}_{i}^{U}$ refers to the mean of response variable for unsupervised models, and $\bar{Y}_{i}^{S} $ is the mean for supervised models.  
	
	\item Construct the log likelihood function, and obtain the maximum likelihood estimator $ \hat{\theta}$ ($ \theta $ refers to ES).
	
	\begin{equation*}\label{logfunc}
	\centering
	\small
	\begin{split}
	& \displaystyle{\sum_{i=1}^{k}} \{X_{i} log[1-\Phi (-\sqrt{\bar{{n}_{i}}}\theta)] + (1-X_{i}) log \Phi (-\sqrt{\bar{{n}_{i}}}\theta) \}
	\end{split}		
	\end{equation*}
	
	where $ k $ is the number of eligible vote entities, $ \Phi $ is the cumulative distribution function, $ \bar{{n}_{i}}$ is average sample size, $\bar{{n}_{i}} = (n1 \times n2)/(n1+n2)$, $n1$ and $n2$ are the sample sizes of unsupervised and supervised experiments respectively. 
	
	\item Compute the 95\% confidence interval for $\hat{\theta}$: [$\theta_{L}$, $\theta_{U}$] .
	
	\begin{center}	
		\centering 
		$\theta_{L} = \hat{\theta} - C_{\alpha/2} \sqrt{Var(\hat{\theta})}$
		
		$\theta_{U} = \hat{\theta} + C_{\alpha/2} \sqrt{Var(\hat{\theta})}$
	\end{center} 
	
	where, $Var(\hat{\theta}) $ is the variance of $ \hat{\theta} $. $ C_{\alpha/2} $ is the two-tailed critical value of the standard normal distribution. In this paper, we use $ C_{\alpha/2}=1.96 $ (central area=0.95, $Z_{\alpha} $=1.96). The variance computation is:
	
	\begin{center}
		$ Var(\hat{\theta}) = \{\displaystyle{\sum_{i=1}^{k}} (
		D_{i}^{1} + D_{i}^{2} - (D_{i}^{1})^2 \frac{(1-2p_{i})}{p_{i}(1-p_{i})}) \}^{-1}$	
	\end{center} 
	
	where,	
	
	$p_{i} = p(\delta,\bar{{n}_{i}}) = 1-\Phi (-\sqrt{\bar{{n}_{i}}}\delta)) $, 
	
	$D_{i}^{1}=	\frac{\partial{p(\delta,\bar{{n}_{i}})}}{\partial{\delta}}
	= \sqrt{\frac{\bar{{n}_{i}}}{2\pi}} exp(-\frac{1}{2}\bar{{n}_{i}} \delta^2 )$, 
	
	$D_{i}^{2}=	\frac{\partial^2{p({\delta},\bar{{n}_{i}})}}{\partial{\delta}^2}
	= -{\frac{\bar{{n}_{i}}*{\delta}}{\sqrt{2\pi}}} exp(-\frac{1}{2}\bar{{n}_{i}} \delta^2 )$.	
	
	$\delta $ is the population standardized mean difference, and it is equal to the estimated $\hat{\theta}$ in Step 4. 
	
\end{enumerate}

\begin{figure*} \centering    
	\subfigure[Vote-counting by study] { \label{fig:vote-paper}     
		\includegraphics[width=0.45\textwidth]{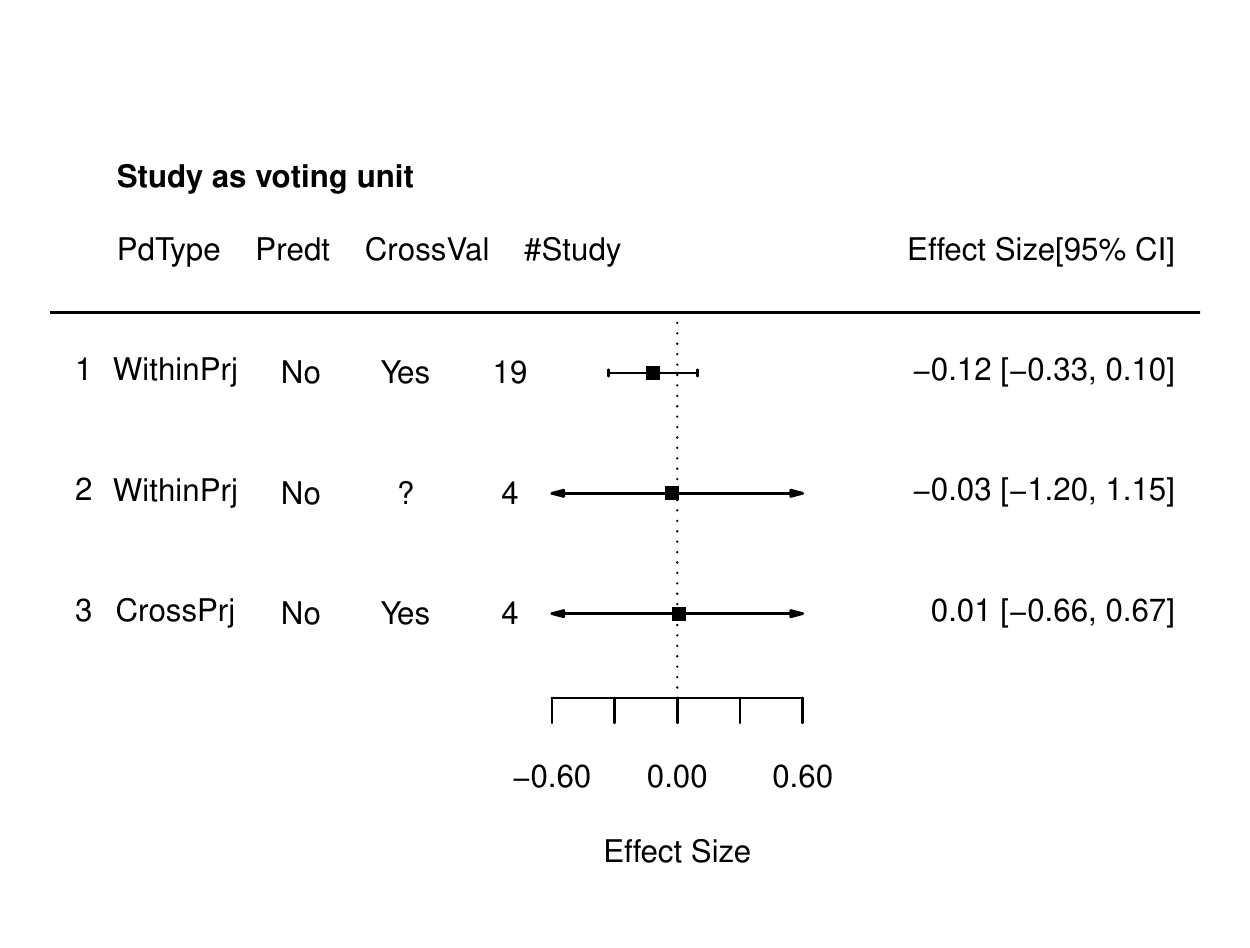}}     
	\subfigure[Vote-counting by project] { \label{fig:vote-prj}     
		\includegraphics[width=0.45\textwidth]{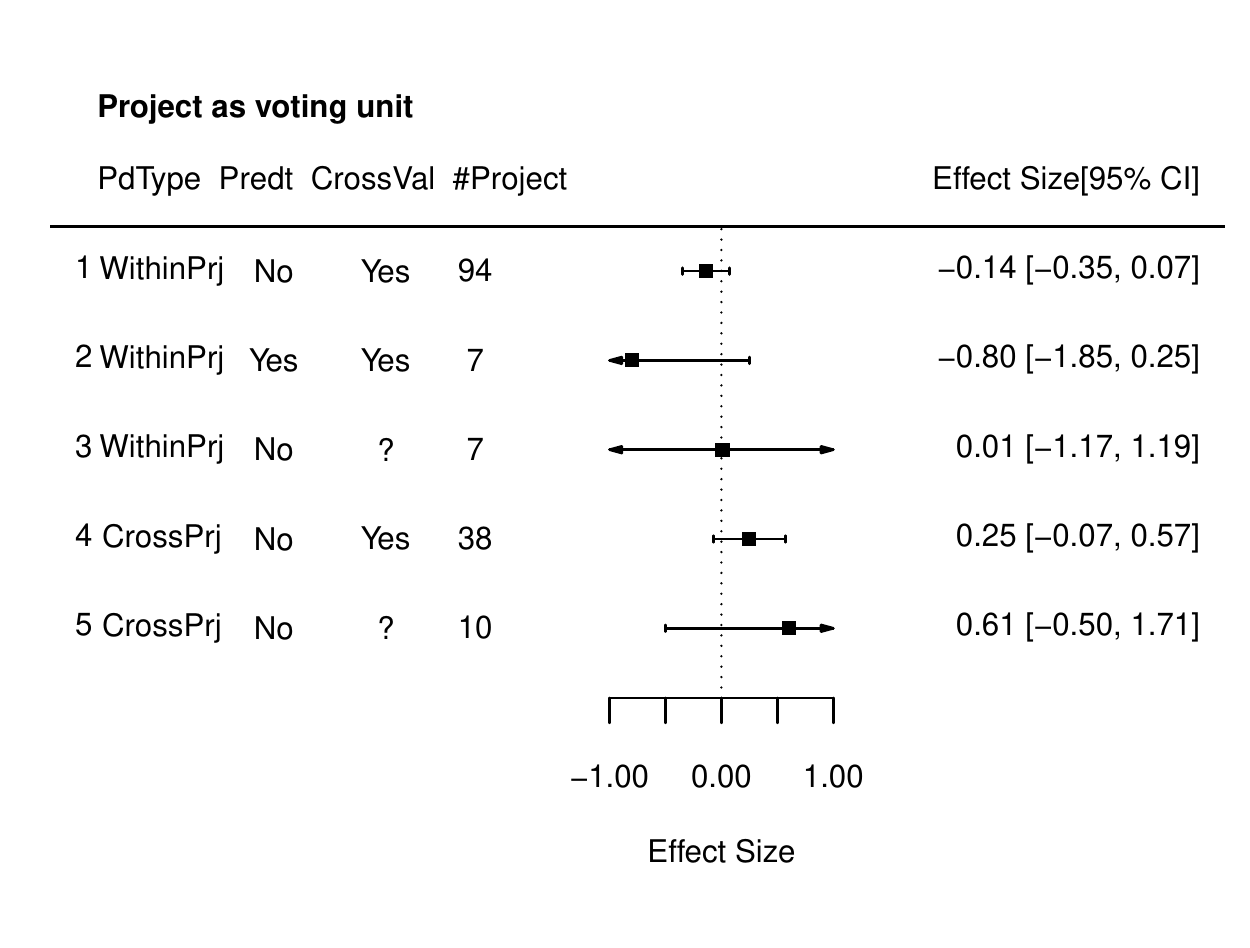} }     
	\caption{Forest plot of UnSDP vs.\ SDP by vote-counting}     
	\label{fig:vote}
	\raggedright{{\footnotesize PdType is prediction type (either within or cross-project), Predt denotes whether the publisher is `predatory' or not, and CrossVal indicates if cross-validation is employed (Yes or ?=unstated) - see Table~\ref{tbl:statClass} for more details. \#Study and \#Project are comparison unit counts.  }}   
\end{figure*}

In terms of voting unit, from our meta-analysis there are 26 studies (papers) or 110 projects that conduct both unsupervised and supervised learning for defect prediction.  If we carry out vote-counting for all combinations of Prediction Type, Predatory and Cross validation, unfortunately there are only sufficient results for 8 conditions, these are listed in bold in Table~\ref{tbl:votecounting}.  This is because the number of eligible study or project results is only zero or one for the other 8 combinations, e.g., if we use project as the voting unit there are no results for cross-project prediction from `predatory' publishers that explicitly use cross-validation.  Another point worth mentioning is that the sum of studies or projects in Table~\ref{tbl:votecounting} or Figure~\ref{fig:vote} is not necessarily 26 or 110. For example: one study may include both within-project and cross-project prediction, it would be counted more than once. 

\begin{table}[ht]
	\caption{Valid combinations for vote-counting} 
	\label{tbl:votecounting}
	\centering
	\begin{tabular}{|r|r|r|r|r|}
		\hline
		Voting & Pred &  Pred & Cross& Count\\ 
		Unit &Type & Pub? & Val? & \\
		\hline
		Study 	& WithinPrj  	& Yes	& Yes 		& 1	\\  
		Study  	& WithinPrj  	& Yes 	& ? 		& 1	\\
		Study  	& WithinPrj  	& No 	& Yes 		& \textbf{19}	\\
		Study  	& WithinPrj  	& No 	& ? 		& \textbf{4}	\\
		Study 	& CrossPrj  	& Yes	& Yes 		& 0	\\  
		Study  	& CrossPrj  	& Yes 	& ? 		& 0	\\
		Study  	& CrossPrj  	& No 	& Yes 		& \textbf{4}	\\
		Study  	& CrossPrj  	& No 	& ? 		& 1	\\
		\hline
		Project & WithinPrj  	& Yes 	& Yes 		& \textbf{7}	\\
		Project & WithinPrj  	& Yes 	& ? 		& 1	\\
		Project & WithinPrj  	& No 	& Yes 		& \textbf{94}\\
		Project & WithinPrj  	& No 	& ? 		& \textbf{7}	\\
		Project & CrossPrj  	& Yes	& Yes 		& 0	\\  
		Project & CrossPrj  	& Yes 	& ? 		& 0	\\
		Project & CrossPrj  	& No 	& Yes 		& \textbf{38}\\
		Project & CrossPrj  	& No 	& ? 		& \textbf{10}\\
		\hline
	\end{tabular}
\end{table}

Figures~\ref{fig:vote-paper} and \ref{fig:vote-prj} show the results of voting by study (paper) and by (software) project respectively. Here, positive values of ES mean UnSDP models perform better than SDP models and negative values the reverse.  The vertical dashed line indicates no difference ($ES=0$). The length of each horizontal line shows the 95\% Confidence Intervals (CI) and the rectangle the estimate of the effect from the relevant set of results.  Note that some of the CI lines (e.g., the second case in Figure~\ref{fig:vote-paper}) end with an arrow, meaning the lower or upper bound of CI exceeds the bound of the legend $[-0.6 ,0.6]$.

Figure~\ref{fig:vote-paper} shows the results of voting by 26 studies including 2052 experiments.  For within-project prediction, although the sign of effect sizes is negative, both of their 95\% CI contain zero, which indicates that it is possible that neither performs better than the other. In other words, UnSDP appears comparable with SDP.  The same explanation also can be applied in Cross-project prediction as the 95\% CI goes across zero.

Figure~\ref{fig:vote-prj} shows the results of voting by 110 projects including 2128 experiments.  These results are broadly similar to voting by paper in that the 95\% CIs straddle zero for all five cases. So this suggests that for both within-project and cross-project prediction UnSDP has potential, since it has data collection advantages over SDP, specifically the need for labelled training data is reduced or removed.

Note that the CIs are widest when there are only a few data points.  It would also seem that the results are most positive to UnSDP when it is unclear that a cross-validation has been employed.  This suggests a certain degree of caution is warranted before claims of superiority are made.

Overall, the results of vote-counting show that UnSDP models are comparable with SDP models, which could make practitioners and researchers more confident to use UnSDP models. However differences in performance are reduced to single votes.  So next we focus on the more informative performance indicator MCC, despite the fact that this reduces the amount of data available.

\subsubsection{Comparison of UnSDP and SDP using MCC} \label{subsec:finemcc}

\begin{figure*}[bh]
	\centering 
	\includegraphics[width=0.95\textwidth]{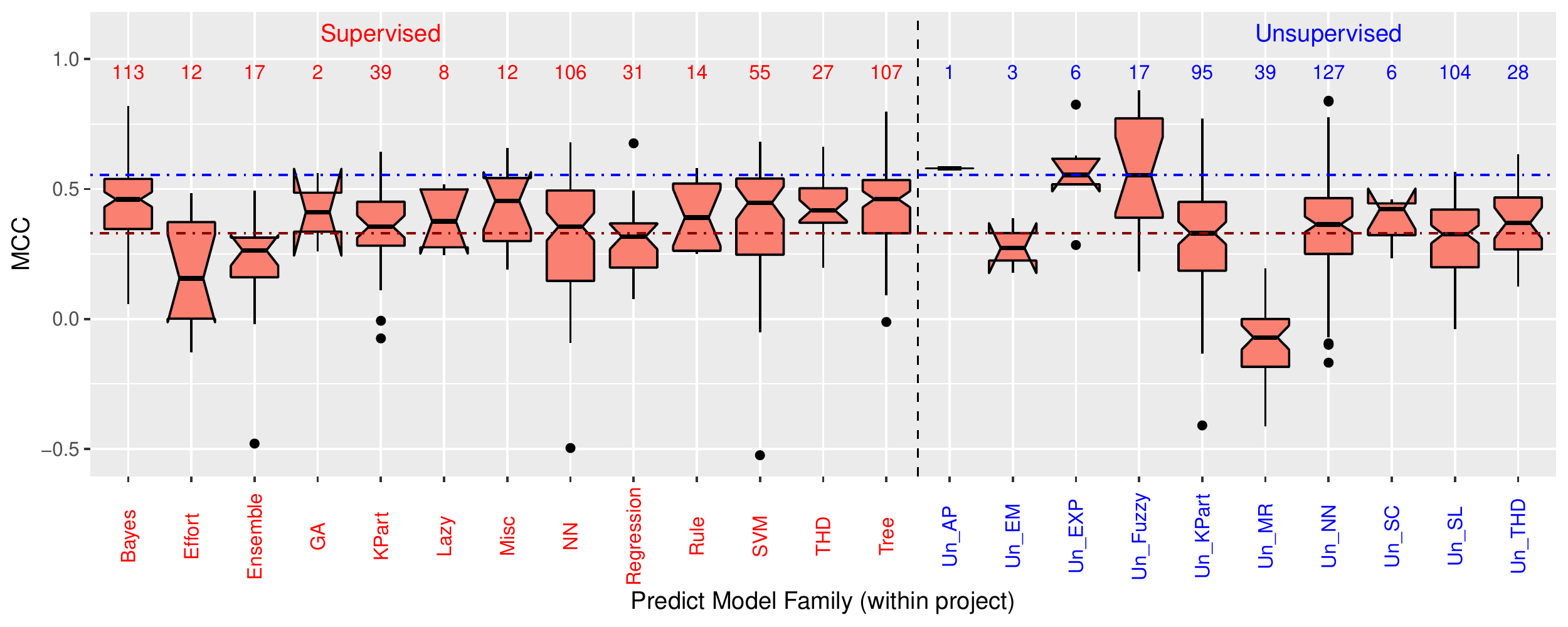}
	\caption{Defect prediction performance (MCC) in within-project prediction by classifier type. N.B.~the numbers indicate the count of experiments (sample size).}
	\label{fig:wp-mcc}
\end{figure*}	

The previous vote-counting comparison does not consider the impact of different prediction performance measures, dataset, etc.  For subsequent analysis we restrict our data to all available, non-predatory results for which MCC is available.  This yields 1178 observations to compare UnSDP and SDP.  N.B.~the MCC results are all recomputed from the reported raw results from the primary studies.

\begin{figure}[ht]
	\centering 
	\includegraphics[width=0.4\textwidth]{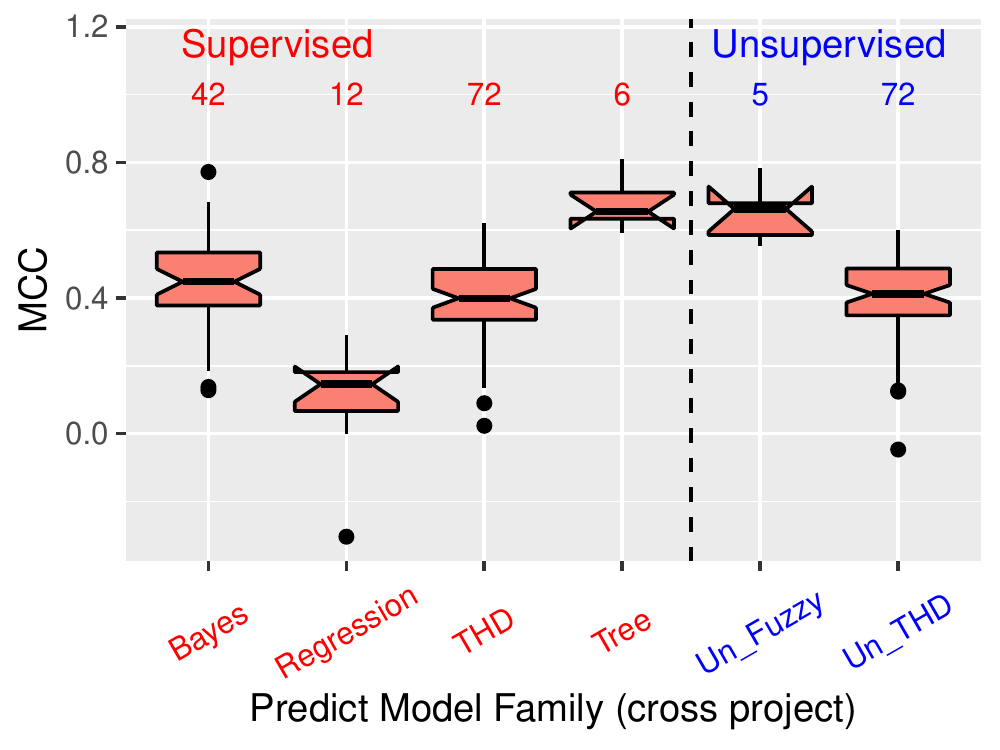}
	\caption{Defect prediction performance (MCC) in cross-project prediction. N.B.~the numbers indicate the count of experiments (sample size).}
	\label{fig:cp-mcc}
\end{figure}

The side by side boxplots in Figure~\ref{fig:wp-mcc} and Figure~\ref{fig:cp-mcc} illustrate within-project and cross-project defect prediction performance respectively.  For more details on the prediction model family refer to Table~\ref{tbl:UnSvSDP} (for unsupervised learners) and Table~\ref{tbl:SvSDP} (for supervised learners).  Although sample size varies considerably among these learners (from 1 to 127), they still illustrate some overall patterns. The boxplot notches indicate the 95\% confidence intervals of the median.  For smaller samples these are wider. Although not a formal test, there is evidence that medians differ if the notches do not overlap \cite{chambers1983graphical}. 

From Figure~\ref{fig:wp-mcc}, we observe that unsupervised models for within-project prediction show greater variance in prediction performance than the supervised ones.  Although it is hard to obtain a definitive overall conclusion which is better, these results are consistent with our vote-counting results, namely the evidence does not differentiate their relative performances to any substantial degree.  

In more detail, and restricting our comments to sample sizes $ >  10$, and therefore more reliable (from Figure~\ref{fig:wp-mcc}) we note the following.

\begin{enumerate}
	\item Un\_Fuzzy (the blue line in Figure~\ref{fig:wp-mcc}) outperforms other SDP and UnSDP model families in that the median of Un\_Fuzzy is higher than that of the other families of learners. However, the CI is wide so it overlaps with other models at least in some contexts.  Among the supervised approaches Bayes appears to be the best model which has also been noted by previous studies \cite{hall2012systematic}.
	\item Traditional unsupervised KPart (partition-based clustering including the widely used KMeans and its variants) family did not perform very well.  The median of KPart (the red line in Figure~\ref{fig:wp-mcc}, 0.33) is lower than the median of all SDP models(0.41). This suggests that researchers should be cautious in using KMeans as a benchmark with which to compare their new UnSDP.
	\item Un\_MR (metric or 1/metric ranking) performs least well.  However, this is based on only a small subset of  the data, since only 39 out of the 234 Un\_MR experimental results use MCC with the majority only reporting Popt or ACC. 
\end{enumerate}

Figure~\ref{fig:cp-mcc} shows the performance of cross-project prediction models.  Un\_Fuzzy again seems to have the best performance.   Presently there are quite limited UnSDP results so further experimentation would be welcome.

We also observe that an unsupervised spectrum clustering models approach (Un\_SC family) was proposed in \citep[P][]{zhang2016cross}, and their model was evaluated with 26 different datasets. It outperformed most of the SDP and other UnSDP models in within-project prediction, and outperformed all others UnSDP and SDP in cross-project prediction.  But unfortunately they only reported AUC, so their model could not be included in Figure~\ref{fig:wp-mcc} and Figure~\ref{fig:cp-mcc}.  This is somewhat frustrating as more complete reporting (of the confusion matrix) would enable integration of their results into meta-analyses.

In summary, UnSDP models are comparable with SDP models both in within-project and cross-project prediction. Compared with most of the SDP models, Un\_Fuzzy, Un\_NN and Un\_SC are potentially the strongest UnSDP approaches. However, we also checked  all  28 primary studies that include SDP models to see whether any parameter tuning was used.  Surprisingly, we found only 3 studies state clearly that tuned SDP models are used in their comparisons, 6 studies use default parameter and 19 studies provide no information about tuning. Therefore, most of SDP models used in comparison might not be best models.  We discuss this potential source of bias in the threats to validity (Section~\ref{sec:Threats}).

\subsubsection{Comparison of UnSDP and SDP using Popt for JiT} \label{subsec:finepopt}

\noindent
Due to incomplete reporting, there are 823 unchecked experimental results that could not be compared with MCC in Section~\ref{subsec:finemcc}. Among these results, AUC and Popt are dominant reported performance measures. Because effort-aware just-in-time (JiT) defect-prone commit prediction differentiates with defect-prone module prediction, effort consideration is its important characteristic. To present a comprehensive comparison of UnSDP and SDP, we carry out the meta-analysis with Popt for JiT prediction in this section. To solid our meta-analysis results, we also compare the unchecked UnSDP and SDP with AUC in Section~\ref{sec:Threats}.

\begin{figure*} \centering    
	\subfigure[Within-project prediction] { \label{fig:wp-popt}     
		\includegraphics[width=0.45\textwidth]{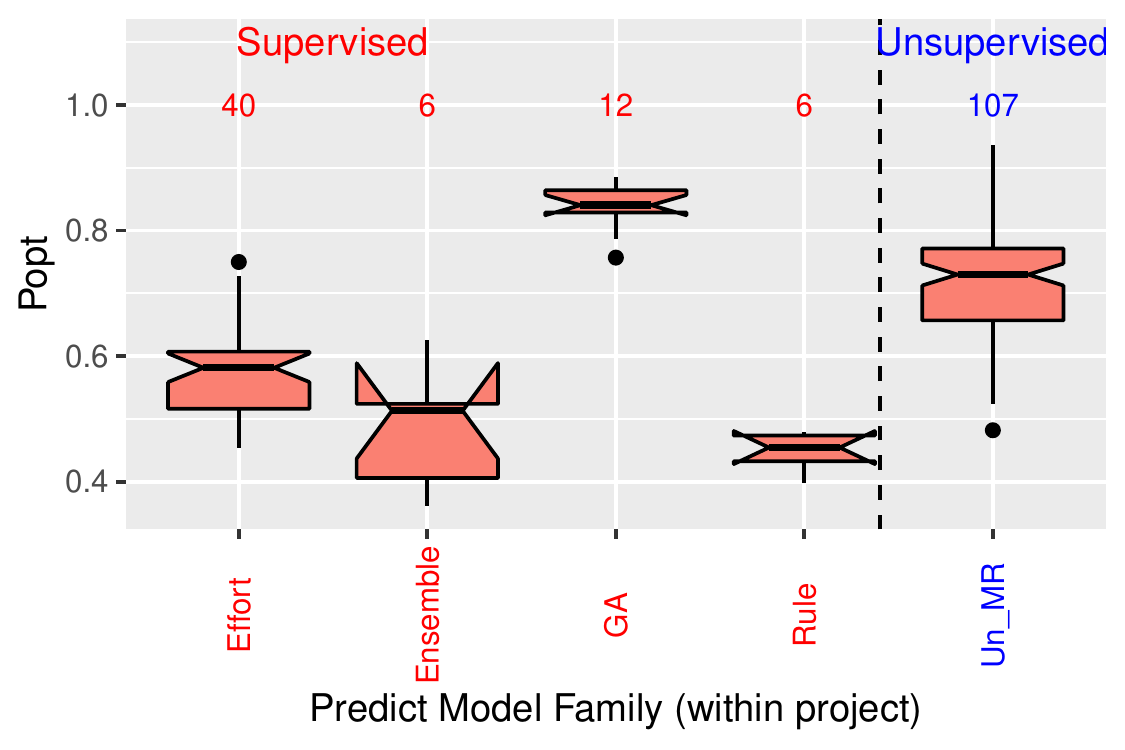}}     
	\subfigure[Cross-project prediction] { \label{fig:cp-popt}     
		\includegraphics[width=0.45\textwidth]{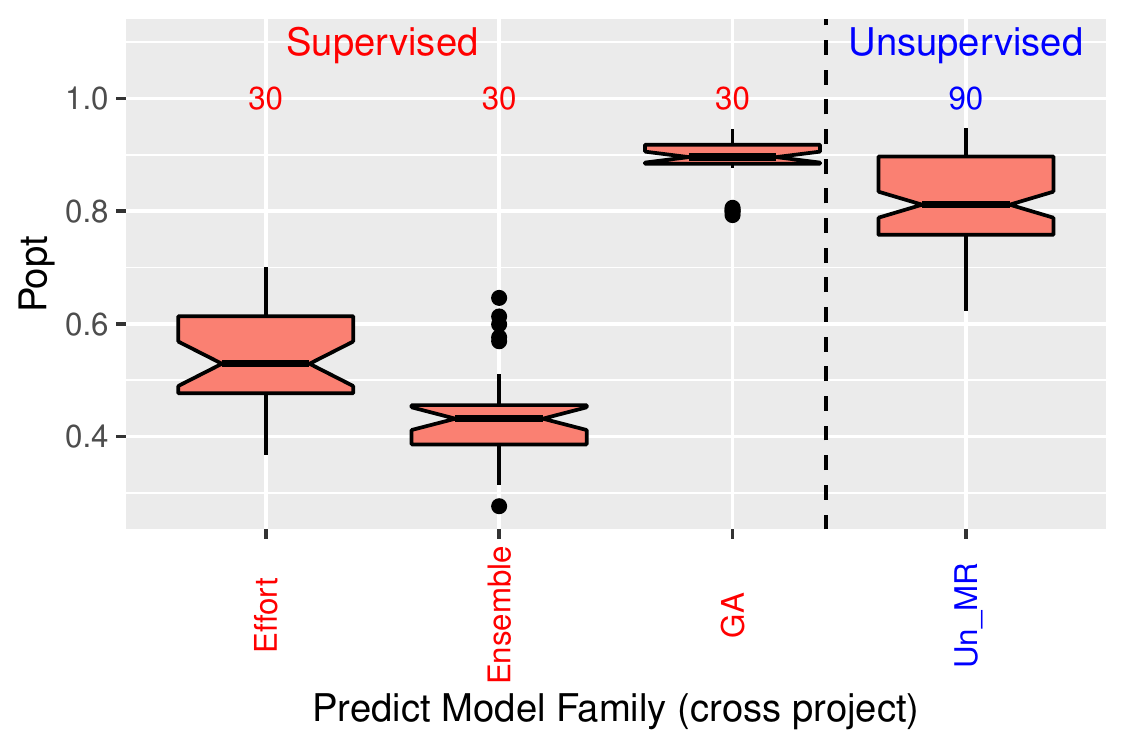} }     
	\caption{Effort-aware just-in-time prediction performance comparison with Popt}     
	\label{fig:popt}
\end{figure*}

Figure~\ref{fig:popt} presents Popt comparison for JiT prediction. It illustrates that SDP model GA (multi-objective optimization based on genetic algorithm) performs better than the only UnSDP model family UN\_MR. However, GA needs to cost more time than UN\_MR for model construction analyzed in \citep[P][]{chen2018multi}. From Figure~\ref{fig:popt}, we can found that UN\_MR is still a good choice for practitioners since it has better prediction performance and less time. In summary, for JiT defect prediction, the comparisons with Popt indicates UnSDP models are comparable with SDP models both in within-project and cross-project prediction.

\subsection{RQ5:Which unsupervised prediction models or model families perform best?}
\label{subsec:UnSDP}

\noindent	
To investigate which models are better among UnSDP models,  we compare them with more detailed subfamily categories. Table~\ref{tbl:UnSDPFamily} lists the mean value of MCC, the corresponding 95\% CI and the number of experiments for all available unsupervised models in within-project defect prediction. Here, we only list the models in which experimental sample size is greater than 1. 

\begin{table}[ht]
	\caption{UnSDP families MCC performance (within-project). Since there is only one experiment in the family of FSubC and AP,  \%95 CI could not be calculated and we removed these two ones from 426 experiments.} 
	\centering
	\label{tbl:UnSDPFamily}
	\footnotesize 
	\begin{threeparttable}
		\begin{tabular}{|r|l|r|r|l|}		\hline
			& MTinyFamily		& MCC 		& \#Exp & \%95 CI \\ 		\hline
			1 & \textbf{FCM} 	& 0.5972 &    11 & [ 0.4357 , 0.7586 ] \\ 
			2 & \textbf{FSOMs} 	& 0.5646 &     5 & [ 0.3309 , 0.7984 ] \\ 
			3 & EXP 			& 0.5600 &     6 & [ 0.3757 , 0.7442 ] \\ 
			4 & KMD 			& 0.4970 &     9 & [ 0.3276 , 0.6665 ] \\ 
			5 & NGas 		    & 0.3949 &    10 & [ 0.1936 , 0.5962 ] \\ 
			6 & SC 				& 0.3799 &     6 & [ 0.2814 , 0.4784 ] \\ 
			7 & THD 			& 0.3740 &    28 & [ 0.3200 , 0.4279 ] \\ 
			8 & ACL 			& 0.3549 &    16 & [ 0.2925 , 0.4172 ] \\ 
			9 & CLA 			& 0.3489 &    23 & [ 0.3014 , 0.3965 ] \\ 
			10 & SOMs 			& 0.3485 &   117 & [ 0.3144 , 0.3826 ] \\ 
			11 & CLAMI 			& 0.3003 &    52 & [ 0.2648 , 0.3358 ] \\ 
			12 & KM 			& 0.2967 &    86 & [ 0.2529 , 0.3406 ] \\ 
			13 & EM 			& 0.2790 &     3 & [ 0.0165 , 0.5414 ] \\ 
			14 & CLAMIPlus $^1$ & 0.2139 &    13 & [ 0.1358 , 0.2920] \\ 
			15 & MR-OneWay $^2$	& -0.0368 &   16 & [ -0.1003 , 0.0268 ] \\ 
			16 & MR 			& -0.1316 &   23 & [ -0.1952 , -0.0681 ] \\ 
			\hline
		\end{tabular}
		\begin{tablenotes}
			\item[1] CLAMIPlus\citep[P][]{yan2017automated} is a variant of CLAMI.
			\item[2] MR-OneWay\citep[P][]{fu2017revisiting} is an improved method based on MR. The same name is also used in Table~\ref{tbl:UnSDPJitMethod}.
		\end{tablenotes}
	\end{threeparttable}
\end{table}

From these results, we can see that FCM and FSOMs learners perform best. Although EXP also performs competitively, we do not recommend it since EXP requires software modules to be classified by an expert manually. Hence we do not consider it to be an effective UnSDP approach.

For JiT defect prediction, we conduct the comparison based on 197 high quality Popt values including 12 UnSDP methods based on MR (metric ranking). Here, we named these methods with MR-$v$ and $v$ is the metric used to rank, e.g. MR-RFC. Also if time-wise cross validation is used in one method, we name it  MR-$v$\_TIME, otherwise it means normal n-fold cross validation is used. Table~\ref{tbl:UnSDPJitMethod} shows MR-CCUM performs best among those unsupervised JiT within-project prediction methods. MR-LT and MR-AGE are followed by MR-CCUM. For 97 cross-project JiT prediction experiments, only MR-CCUM, MR-LT and MR-AGE are included. MR-CCUM is also the best method among them.

\begin{table}[ht]
	\caption{UnSDP methods performance for JiT prediction (within-project)} 
	\centering
	\label{tbl:UnSDPJitMethod}
	\footnotesize 
	\begin{threeparttable}
		\begin{tabular}{|r|l|r|r|l|} \hline
			& MethodName & Popt & \#Exp & \%95CI \\ 
			\hline
			1 & MR-CCUM $^1$ & 0.8930 &     5 & [ 0.8311 , 0.9549 ] \\ 
			2 & MR-CCUM\_TIME & 0.8648 &     6 & [ 0.8217 , 0.908 ] \\ 
			3 & MR-LT$^2$  & 0.7486&    24 & [ 0.7247 , 0.7726 ] \\ 
			4 & MR-AGE$^3$ & 0.7403 &    18 & [ 0.7057 , 0.775 ] \\ 
			5 & MR-LT\_TIME & 0.7120 &     6 & [ 0.6784 , 0.7456 ] \\ 
			6 & MR-NF $^4$ & 0.7065 &     6 & [ 0.6382 , 0.7748 ] \\ 
			7 & MR-AGE\_TIME & 0.7017 &     6 & [ 0.6461 , 0.7572 ] \\ 
			8 & MR-OneWay & 0.6933 &     6 & [ 0.6379 , 0.7487 ] \\ 
			9 & MR-NUC $^5$ & 0.6795 &     6 & [ 0.5863 , 0.7727 ] \\ 
			10 & MR-OneWay\_TIME & 0.6750 &     4 & [ 0.5923 , 0.7577 ] \\ 
			11 & MR-RFC $^6$& 0.6317 &    10 & [ 0.5659 , 0.6975 ] \\ 
			12 & MR-AMC $^7$& 0.6285 &    10 & [ 0.5612 , 0.6958 ] \\ 
			\hline
		\end{tabular}
		\begin{tablenotes}
			\item[1] CCUM : Code churn based unsupervised model. \citep[P][]{yan2017automated}
			\item[2] LT: Lines of code in a file before the current change. \citep[P][]{yang2016effort}
			\item[3] AGE: The average time interval (in days) between the last and
			the change over the files that are touched.\citep[P][]{yang2016effort}
			\item[4] NF:  Number of files touched by the current change. \citep[P][]{yang2016effort}
			\item[5] NUC: Number of unique changes to the modified files. \citep[P][]{yang2016effort}
			\item[6] RFC: Response of a Class. \citep[P][]{yan2017file}
			\item[7] AMC: Average Method Complexity.\citep[P][]{yan2017file}
		\end{tablenotes}
	\end{threeparttable}
\end{table}

\subsection{RQ6: What is the impact of dataset characteristics on predictive performance?}
\label{subsec:Dataset}	

\noindent
Studies (\citep[P][]{yang2016effort} and \citep[P][]{fu2017revisiting}) reveal inconsistent results between predicting defect-prone modules by each project and by all projects as a whole in a just-in-time context. Stimulated by these findings, we also undertake a comparison between UnSDP and SDP models based on individual projects.

Before the comparison of RQ6, we removed two kinds of unsuitable data from Table~\ref{tbl:UnSDPFamily}: (i) EXP related data since expert-based approach is more subjective. (ii) Just-in-time datasets(including BUG, POS, MOZ, PLA, JDT, COL) due to their prominent characteristics compared with defect-prone module prediction dataset.

\begin{figure*}[ht]
	\centering
	\includegraphics[width=0.98\textwidth]{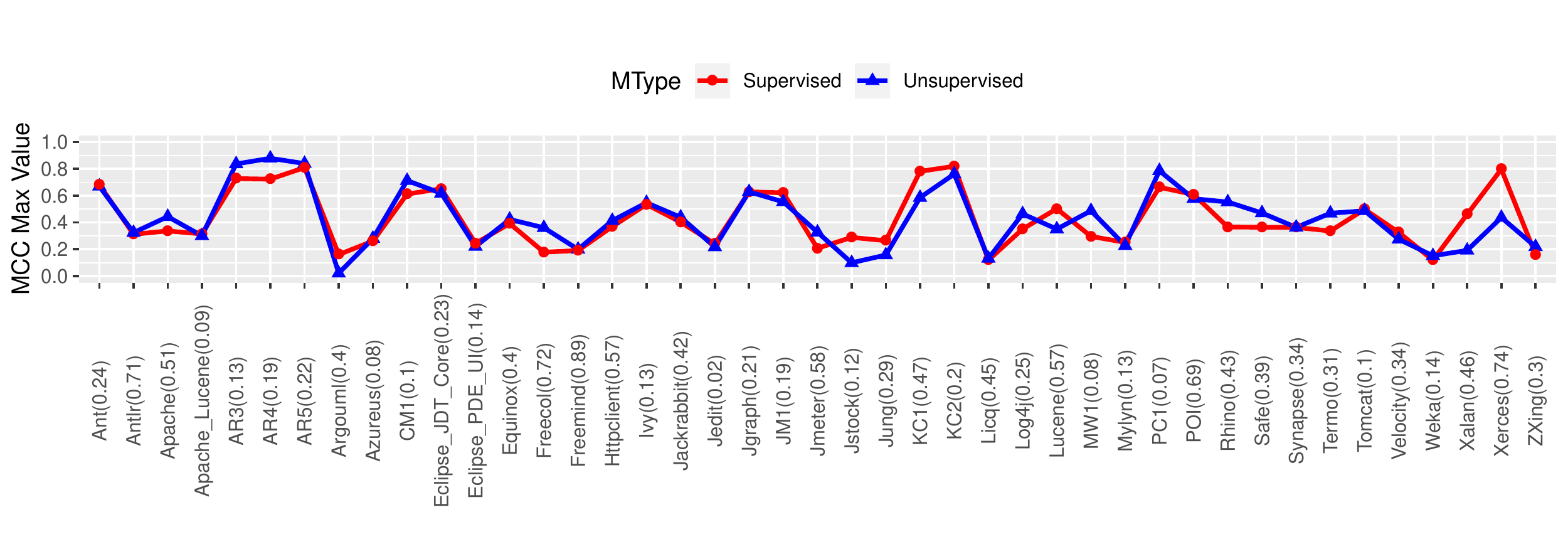}
	\caption{Comparison between UnSDP and SDP by project}
	\label{fig:relation}
\end{figure*}

Figure~\ref{fig:relation} presents the comparison between the \emph{best} UnSDP and SDP models by project. In this figure, horizontal axis label stands for each project and its fault rate. For example, Ant (0.24) denotes the project Ant (Ant1.3, Ant 1.4, etc. are different versions which we unify as Ant). The fault rate 0.24 is average value across all versions.  Figure~\ref{fig:relation} shows that there are similar performance change tendencies for UnSDP and SDP over the different projects. The two groups both have higher prediction performance on some projects (such as Argouml, AR5, Weka, etc.) and lower performance on other projects (such as KC2, ZXing, etc.).  This suggests dataset characteristics have a non-trivial moderating effect on prediction performance.  We analysed the largest two abnormal gaps Xerces and Xalan (Diff $ > $ 0.2) between UnSDP and SDP in Figure~\ref{fig:relation}.  According to the original primary study \citep[P][]{yang2016defect}, the gaps of Xerces and Xalan seem to be caused by their dataset characteristics. 

We investigated one dataset characteristic, namely imbalance (fault rate), but the one-way ANOVA analysis for fault rate and max MCC shows that it is a very weak explanatory factor for MCC (one-way ANOVA F = 0.66, p-value = 0.6261). In summary, we consider the dataset characteristics have an obvious impact on predictive performance. However, it is unclear  which kinds of characteristics are important. We consider this would be well worth exploring in the future.

\section{Threats to Validity} 
\label{sec:Threats}

\noindent	
Threats to \emph{internal validity} relate to our ability to reconstruct a confusion matrix for each experimental result.  Ideally, we would recompute the confusion matrix for all 2456 individual experimental results, however, there are 823 ($~33\%$) results that could not be checked due to incomplete reporting.  Where we could check we found 262 results are problematic, but it may well be that there is an interaction between non-reporting and error-proneness so simply extrapolating from our error-rate findings may not be safe. 

Another threat is our implementation of re-computation with R.  Where possible we compared our partial analysis with the java tool DConfusion \cite{bowes2014dconfusion}.  Our results are consistent with DConfusion.

The third threat is that we only carry out the fine level comparison between UnSDP and SDP with MCC. Although MCC is the best choice for our meta-analysis, to strengthen our analysis, we also conduct an extra analysis with high quality 458 AUC results (Predatory = No and AUC is not NA) . Our comparison results also indicate that Un\_Fuzzy is the most potential approach for within-project prediction. However, the number of experiments is relatively smaller, and more evaluations of Un\_Fuzzy would be recommended. The family of Un\_SC reported only with AUC performs best in cross-project prediction. The analysis results could be found in our Mendeley dataset.

Further, we could only check for consistency errors so it is quite possible that the \emph{actual} rate of experimental analysis errors is greater than we were able to detect.  We have to remark that an overall (knowable) error rate of 11\% across all publication venues does not engender confidence.  We hope a move towards more open science \cite{Muna17} will assist in this regard.

Threats to \emph{external validity} concern the selected experiments and the extent to which the experiments we have located generalise.  By undertaking an explicitly systematic approach \cite{Kitc15} to this review we hope to have included all relevant studies.  These have used a wide range of datasets but we cannot be certain how representative these might be of all possible software defect prediction scenarios.

Another difficulty with the comparison of unsupervised and supervised classification is that the majority of papers appear to focus principally upon novel or innovative uses of UnSDP and that the comparator supervised approaches serve the role of very basic benchmarks.  Consequently it is not always obvious that state of the art supervised algorithms are deployed, nor that much effort has gone into hyper-parameter tuning.  For instance, only 3/28 papers explicitly state they have tuned the SDP models and a further 6/28 indicated defaults were chosen.  Thus there is a danger we are sometimes comparing state of the art UnSDP with off-the-shelf SDP.

A final danger is researcher bias and the tendency to confirm what we already believe to be true.  This is an additional reason why we have made our materials and research processes public$^1$.	

\section{Summary and actionable findings} \label{sec:Conc}

\noindent
UnSDP techniques are attracting more and more researchers and practitioners since labelling information for the training data is not a prerequisite.  We reviewed 49 unsupervised software defect prediction primary studies published between January 2000 to March 2018, which includes 2456 independent experimental results. These 2456 independent results involved 128 software projects (from NASA, PROMISE, ISM, AEEEM, etc.) and 25 prediction model families (172 models). All the UnSDP models and labelling techniques for  clusters are listed in Table~\ref{tbl:UnSvSDP} and Table~\ref{tbl:labelling}. These tables also provide researchers with an indication of the diversity of unsupervised learning techniques used for software defect prediction.

We have carried out two kinds of analysis. 

Firstly, we conducted a bibliometric analysis. We found a growth in research activity in recent years.  We also have found some problems with experimental data quality. We were surprised to find that 25 out of 49 studies did \emph{not} explicitly report whether a cross-validation procedure was used or not, and there were 14 papers (out of 49) that could not be checked for the existence of problematic data due to incomplete reporting.  Therefore, the quality and completeness of reporting should be paid more attention when publishing papers.  Where we were able to assess consistency of results, we found something of the order of 11\% of all results were demonstrably in error due to inconsistencies.  This was an issue in both `predatory' and non-predatory publication venues.

Secondly, we undertook a meta-analysis concerning the performance of UnSDP models.  To compare these results in a reliable way, we recomputed the confusion matrices of the primary studies to obtain consistent performance measures. In our vote-counting analysis, MCC is the main performance measure and others (such as AUC, F1, Popt, etc) are only used when MCC is unavailable.  We found, that UnSDP models are comparable with SDP models both for within-project and cross-project prediction. This indicates UnSDP models may not be as problematic as might be supposed, so we suggest they might be considered in most situations when labelled training data is scarce.  Among the different UnSDP learners, Un\_Fuzzy and Un\_SC appear to have most potential.  Meanwhile, for effort-aware JiT defect prediction, the state-of-the-art complex SDP model GA (multi-objective optimization based on genetic algorithms) performs better than the simple sorting UnSDP model (UN\_MR). However, it less clear whether it has better performance when effort is ignored. Overall, we consider that UN\_MR remains a good choice for practitioners since it has comparable prediction performance to GA model and requires less time. 

We also note that when clustering-based UnSDP approaches are applied, it is a simple and effective way to use the node distribution in clusters (see Table~\ref{tbl:labelling}) to label each cluster as defective or not according to the Pareto principle (sometimes referred to as the 80:20 rule).  

Finally, we found dataset characteristics can have a clear impact on predictive performance, no matter whether UnSDP or SDP models are used. Therefore it will be worth exploring in more depth which of these dataset characteristics are most important. Likewise, the exploration of the interaction between learner and dataset could be fruitful, as it would appear unlikely that a single learning algorithm, be it supervised or unsupervised, will always be optimal.

\section*{Primary Studies} \label{sec:paper}	

\noindent	
Search completed 7th March, 2018.


\makeatletter
\renewcommand\@bibitem[1]{\item\if@filesw \immediate\write\@auxout
	{\string\bibcite{#1}{P\the\value{\@listctr}}}\fi\ignorespaces}
\def\@biblabel#1{[P#1]}
\makeatother

\bibliographystyle{elsarticle-num}

\appendix\

\section {Supervised defect prediction procedure}

\noindent	
For performance comparison, we conclude supervised defect prediction models used in our review studies as Table~\ref{tbl:SvSDP}.

\begin{table*}[ht]
\caption{Supervised Software Defect Prediction Techniques}		
\centering
\footnotesize
\label{tbl:SvSDP}
\begin{tabular}{|p{3.7cm}|p{1.6cm}|p{6.0cm}|p{5cm}|}\hline
	\textbf{Family} & \textbf{Sub Abbr} &\textbf{SubFamily Approaches} &  \textbf{Related Study} \\	\hline
	{\textbf{Bayes}}& NB & 	Na\"ive Bayes &\citep[P][]{c2009clustering}, \citep[P][]{zhong2004analyzing}, \citep[P][]{bishnu2012software}, \citep[P][]{b2011application}, \citep[P][]{singh2014efficient}, \citep[P][]{abaei2015increasing}, \citep[P][]{zhang2016cross}, \citep[P][]{aleem2015benchmarking}, \citep[P][]{abaei2015empirical}, \citep[P][]{fan2017utility}, \citep[P][]{boucher2017predicting}, \citep[P][]{boucher2018software} \\\hline
	\multirow{5}{*}{\textbf{Tree}}& TDT & 	Treedisc Decision Tree &\citep[P][]{zhong2004analyzing} \\\cline{2-4}
	& DT	&Decision tree &\citep[P][]{czibula2016novel},\citep[P][]{aleem2015benchmarking},\citep[P][]{boucher2018software} \\\cline{2-4}
	&DTJ48	&J48 C4.5 &\citep[P][]{zhong2004analyzing},\citep[P][]{yang2016defect},\citep[P][]{aleem2015benchmarking}	 \\\cline{2-4}
	&LMT&	Logistic Model Tree &\citep[P][]{zhang2016cross}\\\cline{2-4}
	&RF		&Random Forest&\citep[P][]{abaei2015increasing},\citep[P][]{zhang2016cross},\citep[P][]{aleem2015benchmarking},\citep[P][]{fu2017revisiting}, \citep[P][]{abaei2015empirical},\citep[P][]{boucher2017predicting}\\\hline
	\multirow{2}{*}{\textbf{SVM}}& SVM & Support Vector Machine &\citep[P][]{z2004unsupervised},\citep[P][]{czibula2016novel},\citep[P][]{aleem2015benchmarking},\citep[P][]{yan2017file},\citep[P][]{yan2017automated},\citep[P][]{fan2017utility},\citep[P][]{boucher2018software} \\\cline{2-4}
	&SMO	&Sequential Minimal Optimization &\citep[P][]{zhong2004analyzing}\\\hline
	\multirow{4}{*}{\textbf{Regression}}& LR & Linear Regression &\citep[P][]{zhong2004analyzing},\citep[P][]{zhang2016cross},\citep[P][]{yang2016defect},\citep[P][]{czibula2016novel},\citep[P][]{n2015clami},\citep[P][]{yan2017file},\citep[P][]{huang2017supervised} \\\cline{2-4}
	&BLR&	Binary Logistic Regression &\citep[P][]{czibula2016novel}\\\cline{2-4}
	&MLR&	Multiple Linear Regression &\citep[P][]{iwata2013error},\citep[P][]{czibula2016novel} \\\hline
	\multirow{3}{*}{\textbf{Lazy}}&kNN& k nearest neighbor &\citep[P][]{zhong2004analyzing}, \citep[P][]{aleem2015benchmarking},\citep[P][]{fu2017revisiting}\\\cline{2-4}
	&LWLS &Locally weighted learning & \citep[P][]{zhong2004analyzing}\\\cline{2-4}
	&CBR&	Case-based reasoning  &\citep[P][]{zhong2004analyzing}\\\hline
	\multirow{6}{*}{\textbf{Rule}}& ONER & One Rule algorithm &\citep[P][]{zhong2004analyzing} \\\cline{2-4}
	&RDR	&Ripple down rule algorithm &\citep[P][]{zhong2004analyzing},\citep[P][]{chen2018multi}\\\cline{2-4}
	&RBM	&Rule-based modeling &\citep[P][]{zhong2004analyzing}	 \\\cline{2-4}
	&JRIP 	&Repeated incremental pruning& \citep[P][]{zhong2004analyzing}\\\cline{2-4}
	&DTABLE	&Decision table		&\citep[P][]{zhong2004analyzing},\citep[P][]{boucher2018software} \\\cline{2-4}
	&ADT	&Alternating decision table	&\citep[P][]{zhong2004analyzing}\\\cline{2-4}
	&PART	&Rules from partial decision trees	&\citep[P][]{zhong2004analyzing}\\\hline
	\multirow{6}{*}{\textbf{Neural Network (NN)}}& ANN & Artificial neural networks &\citep[P][]{zhong2004analyzing},\citep[P][]{czibula2016novel},\citep[P][]{iwata2012clustering},\citep[P][]{abaei2015empirical},\citep[P][]{fan2017utility},\citep[P][]{boucher2017predicting},\citep[P][]{boucher2018software}   \\\cline{2-4}
	&CCN	&Cascade Correlation Networks & \citep[P][]{czibula2016novel}\\\cline{2-4}
	&GMDHN &	GMDH Network			 & \citep[P][]{czibula2016novel}\\\cline{2-4}
	&RBFN	&Radial basis function network &\citep[P][]{aleem2015benchmarking}, \citep[P][]{yan2017file}\\\cline{2-4}
	&MLP	&Multilayer perceptron&\citep[P][]{singh2016comparative},\citep[P][]{aleem2015benchmarking},\citep[P][]{yan2017file}\\\hline
	\multirow{3}{*}{\textbf{Ensemble}}& LBOOST & LogitBoost classifier &\citep[P][]{zhong2004analyzing}, \citep[P][]{yan2017file},\citep[P][]{singh2017classification} \\\cline{2-4}
	&ABOOST	&AdaBoost classifier	&\citep[P][]{zhong2004analyzing},\citep[P][]{aleem2015benchmarking},\citep[P][]{chen2018multi}\\\cline{2-4}
	&BAG	&Bagging classifier		&\citep[P][]{zhong2004analyzing},\citep[P][]{aleem2015benchmarking} \\\hline
	{\textbf{{\footnotesize Genetic Algorithm (GA)}}}
	& GA	 & Genetic algorithms (evolutionary algorithms) &\citep[P][]{zhong2004analyzing},\citep[P][]{czibula2016novel},\citep[P][]{chen2018multi}\\\hline
	\multirow{2}{*}{\textbf{{\footnotesize Effort-aware}}}& EALR & Effort aware&\citep[P][]{yang2016effort},\citep[P][]{fu2017revisiting},\citep[P][]{liu2017code}, \citep[P][]{huang2017supervised},\citep[P][]{chen2018multi} \\\cline{2-4}
	&TLEL	&Two-layer ensemble learning &\citep[P][]{liu2017code} \\\hline
	\multirow{7}{*}{\textbf{Misc}}& RSET & Rough sets-based classifier &\citep[P][]{zhong2004analyzing} \\\cline{2-4}
	&MCOST	&	MetaCost classifier 	&\citep[P][]{zhong2004analyzing}  	\\\cline{2-4}
	&GeneEP	&Gene Expression Programming& \citep[P][]{czibula2016novel} \\\cline{2-4}
	&HPIPES	&Hyperpipes algorithm	&\citep[P][]{zhong2004analyzing}\\\cline{2-4}
	&LDA	&Linear Discriminant Analysis&\citep[P][]{bishnu2012software},\citep[P][]{b2011application} \\\cline{2-4}
	&{\footnotesize SOMs-MLR}&Hybrid: SOMs and Multiple Logic Regression &\citep[P][]{iwata2013error} \\\cline{2-4}
	&STHD	&Threshold determined by supervised method &\citep[P][]{boucher2018software} \\\cline{2-4}
	\hline
	
\end{tabular}
\end{table*}	

\section*{Acknowledgements} \label{sec:Ack}
\noindent
We would like to thank the editors and the anonymous reviewers for their insightful comments and suggestions. We also wish to acknowledge the use of the DConfusion tool developed by David Bowes and David Gray whilst at the University of Hertfordshire.
This work was supported by the National Key Basic Research Program of China [2018YFB1004401]; the National Natural Science Foundation of China [61972317, 61402370].

\section*{References}

\begin{thebibliography}{10}
\bibitem{catal2009software}
C.~Catal, U.~Sevim, B.~Diri, Software fault prediction of unlabeled program
modules, in: Proceedings of the world congress on engineering, Vol.~1, 2009,
pp. 1--6.

\bibitem{c2009clustering}
C.~Catal, U.~Sevim, B.~Diri, Clustering and metrics thresholds based software
fault prediction of unlabeled program modules, in: Information Technology:
New Generations, 2009. ITNG'09. Sixth International Conference on, IEEE,
2009, pp. 199--204.

\bibitem{zhong2004analyzing}
S.~Zhong, T.~M. Khoshgoftaar, N.~Seliya, Analyzing software measurement data
with clustering techniques, IEEE Intelligent Systems 19~(2) (2004) 20--27.

\bibitem{z2004unsupervised}
S.~Zhong, T.~M. Khoshgoftaar, N.~Seliya, Unsupervised learning for expert-based
software quality estimation., in: HASE, 2004, pp. 149--155.

\bibitem{bishnu2012software}
P.~S. Bishnu, V.~Bhattacherjee, Software fault prediction using quad tree-based
k-means clustering algorithm, IEEE Transactions on knowledge and data
engineering 24~(6) (2012) 1146--1150.

\bibitem{b2011application}
P.~S. Bishnu, V.~Bhattacherjee, Application of k-medoids with kd-tree for
software fault prediction, ACM SIGSOFT Software Engineering Notes 36~(2)
(2011) 1--6.

\bibitem{singh2016comparative}
S.~Singh, R.~Singla, Comparative performance of fault-prone prediction classes
with k-means clustering and mlp, in: Proceedings of the Second International
Conference on Information and Communication Technology for Competitive
Strategies, ACM, 2016, p.~65.

\bibitem{varade2013hyper}
S.~Varade, M.~Ingle, Hyper-quad-tree based k-means clustering algorithm for
fault prediction, International Journal of Computer Applications 76~(5).

\bibitem{coelho2014applying}
R.~A. Coelho, F.~dos RN~Guimar{\~a}es, A.~A. Esmin, Applying swarm ensemble
clustering technique for fault prediction using software metrics, in: Machine
Learning and Applications (ICMLA), 2014 13th International Conference on,
IEEE, 2014, pp. 356--361.

\bibitem{chug2013software}
A.~Chug, S.~Dhall, Software defect prediction using supervised learning
algorithm and unsupervised learning algorithm, in: Confluence 2013: The Next
Generation Information Technology Summit (4th International Conference), IET,
2013, pp. 173--179.

\bibitem{gupta2013estimating}
D.~Gupta, V.~K. Goyal, H.~Mittal, Estimating of software quality with
clustering techniques, in: Advanced Computing and Communication Technologies
(ACCT), 2013 Third International Conference on, IEEE, 2013, pp. 20--27.

\bibitem{yang2008software}
B.~Yang, Q.~Yin, S.~Xu, P.~Guo, Software quality prediction using affinity
propagation algorithm, in: Neural Networks, 2008. IJCNN 2008.(IEEE World
Congress on Computational Intelligence). IEEE International Joint Conference
on, IEEE, 2008, pp. 1891--1896.

\bibitem{yang2006software}
B.~Yang, X.~Zheng, P.~Guo, Software metrics data clustering for quality
prediction, Computational Intelligence (2006) 959--964.

\bibitem{kaur2014comparative}
D.~Kaur, A comparative study of various distance measures for software fault
prediction, arXiv preprint arXiv:1411.7474.

\bibitem{sandhu2010k}
P.~S. Sandhu, J.~Singh, V.~Gupta, M.~Kaur, S.~Manhas, R.~Sidhu, A k-means based
clustering approach for finding faulty modules in open source software
systems, World academy of science, Engineering and technology 72 (2010)
654--658.

\bibitem{singh2014efficient}
P.~Singh, S.~Verma, An efficient software fault prediction model using cluster
based classification, International Journal of Applied Information Systems
(IJAIS) 7~(3) (2014) 35--41.

\bibitem{kaur2010clustering}
D.~Kaur, A.~Kaur, S.~Gulati, M.~Aggarwal, A clustering algorithm for software
fault prediction, in: Computer and Communication Technology (ICCCT), 2010
International Conference on, IEEE, 2010, pp. 603--607.

\bibitem{abaei2015increasing}
G.~Abaei, A.~Selamat, Increasing the accuracy of software fault prediction
using majority ranking fuzzy clustering, in: Software Engineering, Artificial
Intelligence, Networking and Parallel/Distributed Computing, Springer, 2015,
pp. 179--193.

\bibitem{catal2010metrics}
C.~Catal, U.~Sevim, B.~Diri, Metrics-driven software quality prediction without
prior fault data, in: Electronic Engineering and Computing Technology,
Springer, 2010, pp. 189--199.

\bibitem{park2014software}
M.~Park, E.~Hong, Software fault prediction model using clustering algorithms
determining the number of clusters automatically, International Journal of
Software Engineering \& Its Applications 8~(7) (2014) 199--204.

\bibitem{guo2000software}
P.~Guo, M.~R. Lyu, Software quality prediction using mixture models with em
algorithm, in: Quality Software, 2000. Proceedings. First Asia-Pacific
Conference on, IEEE, 2000, pp. 69--78.

\bibitem{sandhu2010density}
P.~S. Sandhu, M.~Kaur, A.~Kaur, A density based clustering approach for early
detection of fault prone modules, in: Electronics and Information Engineering
(ICEIE), 2010 International Conference On, Vol.~2, IEEE, 2010, pp. 525--530.

\bibitem{yuan2000application}
X.~Yuan, T.~M. Khoshgoftaar, E.~B. Allen, K.~Ganesan, An application of fuzzy
clustering to software quality prediction, in: Application-Specific Systems
and Software Engineering Technology, 2000. Proceedings. 3rd IEEE Symposium
on, IEEE, 2000, pp. 85--90.

\bibitem{sidhu2010subtractive}
R.~S. Sidhu, S.~Khullar, P.~S. Sandhu, R.~Bedi, K.~Kaur, A subtractive
clustering based approach for early prediction of fault proneness in software
modules, World Academy of Science, Engineering and Technology 4~(7)
1165--1169.

\bibitem{zhang2016cross}
F.~Zhang, Q.~Zheng, Y.~Zou, A.~E. Hassan, Cross-project defect prediction using
a connectivity-based unsupervised classifier, in: Proceedings of the 38th
International Conference on Software Engineering, ACM, 2016, pp. 309--320.

\bibitem{yang2016defect}
J.~Yang, H.~Qian, Defect prediction on unlabeled datasets by using unsupervised
clustering, in: High Performance Computing and Communications; IEEE 14th
International Conference on Smart City; IEEE 2nd International Conference on
Data Science and Systems (HPCC/SmartCity/DSS), 2016 IEEE 18th International
Conference on, IEEE, 2016, pp. 465--472.

\bibitem{yang2008Metrics}
B.~Yang, X.~Chen, S.~Xu, P.~Guo, Software metrics analysis with genetic
algorithm and affinity propagation clustering, in: DMIN, 2008, pp. 590--596.

\bibitem{iwata2013error}
K.~Iwata, T.~Nakashima, Y.~Anan, N.~Ishii, Error prediction methods for
embedded software development using hybrid models of self-organizing maps and
multiple regression analyses, in: Software Engineering, Artificial
Intelligence, Networking and Parallel/Distributed Computing 2013, Springer,
2013, pp. 185--200.

\bibitem{abaei2013fault}
G.~Abaei, Z.~Rezaei, A.~Selamat, Fault prediction by utilizing self-organizing
map and threshold, in: Control System, Computing and Engineering (ICCSCE),
2013 IEEE International Conference on, IEEE, 2013, pp. 465--470.

\bibitem{marian2015software}
Z.~Marian, G.~Czibula, G.~Czibula, S.~Sotoc, Software defect detection using
self-organizing maps, Studia Universitatis Babes-Bolyai, Informatica 60~(2)
(2015) 55--69.

\bibitem{czibula2016novel}
I.-G. Czibula, G.~Czibula, Z.~Marian, V.-S. Ionescu, A novel approach using
fuzzy self-organizing maps for detecting software faults, Studies in
Informatics and Control 25~(2) (2016) 207--216.

\bibitem{iwata2012clustering}
K.~Iwata, T.~Nakashima, Y.~Anan, N.~Ishii, Clustering and analyzing embedded
software development projects data using self-organizing maps, Software
Engineering Research, Management and Applications 2012 (2012) 47--59.

\bibitem{aleem2015benchmarking}
S.~Aleem, L.~Capretz, F.~Ahmed, Benchmarking machine learning techniques for
software defect detection, Int. J. Softw. Eng. Appl 6~(3).

\bibitem{n2015clami}
J.~Nam, S.~Kim, Clami: Defect prediction on unlabeled datasets, in: Automated
Software Engineering (ASE), 2015 30th IEEE/ACM International Conference on,
IEEE, 2015, pp. 452--463.

\bibitem{catal2013fault}
C.~Catal, B.~Diri, A fault detection strategy for software projects,
Tehni{\v{c}}ki vjesnik 20~(1) (2013) 1--7.

\bibitem{yang2016effort}
Y.~Yang, Y.~Zhou, J.~Liu, Y.~Zhao, H.~Lu, L.~Xu, B.~Xu, H.~Leung, Effort-aware
just-in-time defect prediction: simple unsupervised models could be better
than supervised models, in: Proceedings of the 2016 24th ACM SIGSOFT
International Symposium on Foundations of Software Engineering, ACM, 2016,
pp. 157--168.

\bibitem{fu2017revisiting}
W.~Fu, T.~Menzies, Revisiting unsupervised learning for defect prediction, in:
Proceedings of the 2017 11th Joint Meeting on Foundations of Software
Engineering, ACM, 2017, pp. 72--83.

\bibitem{abaei2015empirical}
G.~Abaei, A.~Selamat, H.~Fujita, An empirical study based on semi-supervised
hybrid self-organizing map for software fault prediction, Knowledge-Based
Systems 74 (2015) 28--39.

\bibitem{liu2017code}
J.~Liu, Y.~Zhou, Y.~Yang, H.~Lu, B.~Xu, Code churn: A neglected metric in
effort-aware just-in-time defect prediction, in: Empirical Software
Engineering and Measurement (ESEM), 2017 ACM/IEEE International Symposium on,
IEEE, 2017, pp. 11--19.

\bibitem{yan2017file}
M.~Yan, Y.~Fang, D.~Lo, X.~Xia, X.~Zhang, File-level defect prediction:
Unsupervised vs. supervised models, in: Empirical Software Engineering and
Measurement (ESEM), 2017 ACM/IEEE International Symposium on, IEEE, 2017, pp.
344--353.

\bibitem{fan2017utility}
Y.~Fan, C.~Lv, X.~Zhang, G.~Zhou, Y.~Zhou, The utility challenge of
privacy-preserving data-sharing in cross-company defect prediction: An
empirical study of the cliff\&morph algorithm, in: Software Maintenance and
Evolution (ICSME), 2017 IEEE International Conference on, IEEE, 2017, pp.
80--90.

\bibitem{huang2017supervised}
Q.~Huang, X.~Xia, D.~Lo, Supervised vs unsupervised models: A holistic look at
effort-aware just-in-time defect prediction, in: Software Maintenance and
Evolution (ICSME), 2017 IEEE International Conference on, IEEE, 2017, pp.
159--170.

\bibitem{singh2017classification}
S.~Singh, R.~Singla, Classification of defective modules using object-oriented
metrics, International Journal of Intelligent Systems Technologies and
Applications 16~(1) (2017) 1--13.

\bibitem{chang2017novel}
R.~Chang, X.~Shen, B.~Wang, Q.~Xu, A novel method for software defect
prediction in the context of big data, in: Big Data Analysis (ICBDA), 2017
IEEE 2nd International Conference on, IEEE, 2017, pp. 100--104.

\bibitem{boucher2017predicting}
A.~Boucher, M.~Badri, Predicting fault-prone classes in object-oriented
software: An adaptation of an unsupervised hybrid som algorithm, in: Software
Quality, Reliability and Security (QRS), 2017 IEEE International Conference
on, IEEE, 2017, pp. 306--317.

\bibitem{yan2017automated}
M.~Yan, X.~Zhang, C.~Liu, L.~Xu, M.~Yang, D.~Yang, Automated change-prone class
prediction on unlabeled dataset using unsupervised method, Information and
Software Technology 92 (2017) 1--16.

\bibitem{boucher2018software}
A.~Boucher, M.~Badri, Software metrics thresholds calculation techniques to
predict fault-proneness: An empirical comparison, Information and Software
Technology 96 (2018) 38--67.

\bibitem{sasidharan2014hyper}
R.~Sasidharan, P.~Sriram, Hyper-quadtree-based k-means algorithm for software
fault prediction (2014) 107--118.

\bibitem{chen2018multi}
X.~Chen, Y.~Zhao, Q.~Wang, Z.~Yuan, Multi: Multi-objective effort-aware
just-in-time software defect prediction, Information and Software Technology
93 (2018) 1--13.
\end{thebibliography}

\begin{thebibliography}{10}
\bibitem{fenton1999critique}
N.~E. Fenton, M.~Neil, A critique of software defect prediction models, IEEE
Transactions on software engineering 25~(5) (1999) 675--689.

\bibitem{catal2009systematic}
C.~Catal, B.~Diri, A systematic review of software fault prediction studies,
Expert systems with applications 36~(4) (2009) 7346--7354.

\bibitem{hall2012systematic}
T.~Hall, S.~Beecham, D.~Bowes, D.~Gray, S.~Counsell, A systematic literature
review on fault prediction performance in software engineering, IEEE
Transactions on Software Engineering 38~(6) (2012) 1276--1304.

\bibitem{nam2015clami}
J.~Nam, S.~Kim, Clami: Defect prediction on unlabeled datasets (t), in:
Automated Software Engineering (ASE), 2015 30th IEEE/ACM International
Conference on, IEEE, 2015, pp. 452--463.

\bibitem{Kitc15}
B.~Kitchenham, D.~Budgen, P.~Brereton, Evidence-based Software Engineering and
Systematic Reviews, CRC Press, 2015.

\bibitem{yan2016self}
M.~Yan, M.~Yang, C.~Liu, X.~Zhang, Self-learning change-prone class
prediction., in: SEKE, 2016, pp. 134--140.

\bibitem{boucher2016using}
A.~Boucher, M.~Badri, Using software metrics thresholds to predict fault-prone
classes in object-oriented software, in: 2016 4th Intl Conf on Applied
Computing and Information Technology/3rd Intl Conf on Computational
Science/Intelligence and Applied Informatics/1st Intl Conf on Big Data, Cloud
Computing, Data Science \& Engineering (ACIT-CSII-BCD), IEEE, 2016, pp.
169--176.

\bibitem{Li2020Data}
N.~Li, M.~Shepperd, Y.~Guo, A Systematic Review of Unsupervised Defect Prediction Dataset, Mendeley Data,v1. \url{http://dx.doi.org/10.17632/h24ctmyx73.1}

\bibitem{Powe11}
D.~Powers, Evaluation: from precision, recall and {F}-measure to {ROC},
informedness, markedness and correlation, Journal of Machine Learning
Technologies 2~(1) (2011) 37--63.

\bibitem{shepperd2014researcher}
M.~Shepperd, D.~Bowes, T.~Hall, Researcher bias: The use of machine learning in
software defect prediction, IEEE Transactions on Software Engineering 40~(6)
(2014) 603--616.

\bibitem{bowes2014dconfusion}
D.~Bowes, T.~Hall, D.~Gray, {DC}onfusion: a technique to allow cross study
performance evaluation of fault prediction studies, Automated Software
Engineering 21~(2) (2014) 287--313.

\bibitem{baldi2000assessing}
P.~Baldi, S.~Brunak, Y.~Chauvin, C.~A. Andersen, H.~Nielsen, Assessing the
accuracy of prediction algorithms for classification: an overview,
Bioinformatics 16~(5) (2000) 412--424.

\bibitem{fawcett2006introduction}
T.~Fawcett, An introduction to roc analysis, Pattern recognition letters 27~(8)
(2006) 861--874.

\bibitem{kamei2013large}
Y.~Kamei, E.~Shihab, B.~Adams, A.~E. Hassan, A.~Mockus, A.~Sinha, N.~Ubayashi,
A large-scale empirical study of just-in-time quality assurance, IEEE
Transactions on Software Engineering 39~(6) (2013) 757--773.

\bibitem{Hand09}
D.~Hand, Measuring classifier performance: a coherent alternative to the area
under the {ROC} curve, Machine Learning 77~(1) (2009) 103--123.

\bibitem{Flac15}
P.~Flach, M.~Kull, Precision-recall-gain curves: {PR} analysis done right, in:
Advances in Neural Information Processing Systems, 2015, pp. 838--846.

\bibitem{song2018comprehensive}
Q.~Song, Y.~Guo, M.~Shepperd, A comprehensive investigation of the role of
imbalanced learning for software defect prediction, IEEE Transactions on
Software Engineering 45~(12) (2018) 1253--1269.

\bibitem{perlin2017predatory}
M.~S. Perlin, T.~Imasato, D.~Borenstein, Is predatory publishing a real threat?
evidence from a large database study, Scientometrics (2017) 1--19.

\bibitem{Berg15}
M.~Berger, J.~Cirasella, Beyond {B}eall's list: {B}etter understanding
predatory publishers, College \& Research Libraries News 76~(3) (2015)
132--135.

\bibitem{kohavi1995study}
R.~Kohavi, et~al., A study of cross-validation and bootstrap for accuracy
estimation and model selection, in: Ijcai, Vol.~14, Montreal, Canada, 1995,
pp. 1137--1145.

\bibitem{grissom2012effect}
R.~J. Grissom, J.~J. Kim, Effect sizes for research: Univariate and
multivariate applications, Routledge, 2012.

\bibitem{Mich94}
D.~Michie, D.~Spiegelhalter, C.~Taylor, Machine learning, neural and
statistical classification, Ellis Horwood Series in Artifical Intelligence,
Ellis Horwood, Chichester, Sussex, UK, 1994.

\bibitem{Bush09}
B.~Bushman, M.~Wang, Vote-counting procedures in meta-analysis, Russell Sage
Foundation, New York, NY, US, 2009, pp. 207--220.

\bibitem{Most96}
F.~Mosteller, G.~Colditz, Understanding research synthesis (meta-analysis),
Annual Review of Public Health 17~(1) (1996) 1--23.

\bibitem{Hedges1985statistical}
L.~V. Hedges, I.~Olkin, Statistical methods for meta-analysis, Academic Press,
1985.

\bibitem{chambers1983graphical}
J.~M. Chambers, Graphical methods for data analysis, Wadsworth International
Group, 1983.

\bibitem{Muna17}
M.~Munaf{\`o}, B.~Nosek, D.~Bishop, K.~Button, C.~Chambers, N.~du~Sert,
U.~Simonsohn, E.~Wagenmakers, J.~Ware, J.~Ioannidis, A manifesto for
reproducible science, Nature Human Behaviour 1~(1) (2017) 0021.

	
\end{thebibliography}

\makeatletter
\renewcommand\@bibitem[1]{\item\if@filesw \immediate\write\@auxout
	{\string\bibcite{#1}{\the\value{\@listctr}}}\fi\ignorespaces}
\def\@biblabel#1{[#1]}
\makeatother

\bibliographystyle{elsarticle-num}

\end{document}